\documentclass[twoside,12pt]{article}
\usepackage[dvipsnames,table,xcdraw]{xcolor}
\definecolor{awesome}{rgb}{1.0, 0.13, 0.32}
\definecolor{dodgerblue}{rgb}{0.12, 0.56, 1.0}
\definecolor{trueblue}{rgb}{0.0, 0.45, 0.81}
\definecolor{amaranth}{rgb}{0.9, 0.17, 0.31}
\usepackage[T1]{fontenc}

\usepackage{comment}
\usepackage[top=1in, bottom=1in, left=1.5in, right=1in]{geometry}
\usepackage{rotating}
\usepackage{amsmath}
\usepackage{amsfonts}
\usepackage{algorithm}
\usepackage{algorithmic}
\usepackage{multirow}
\usepackage{colortbl}
\usepackage{color}
\usepackage{epigraph}
\usepackage{graphicx}
\usepackage{caption}
\usepackage{subcaption}
\usepackage{tabularx}
\usepackage{float}
\usepackage{longtable}
\usepackage{pdfpages}
\usepackage{pdflscape}
\usepackage[acronym,toc]{glossaries}
\usepackage{setspace}
\usepackage{pifont}
\usepackage{relsize}
\usepackage{bm}
\usepackage{amssymb}
\usepackage{siunitx}
\usepackage{url}
\usepackage{setspace}
\usepackage{amsmath}
\usepackage{amsthm}
\usepackage{amssymb}
\usepackage{mathrsfs}
\usepackage{mathtools}
\usepackage{rotating}
\usepackage{adjustbox}
\usepackage{upgreek}
\usepackage{tikz}
\usetikzlibrary{arrows,calc,shapes,decorations.pathreplacing}
\usepackage{pgfplots}

\theoremstyle{plain}
 
\theoremstyle{definition}
\usepackage{titlesec}
\usepackage{pdfpages}
\usepackage{changepage}
\usepackage{realboxes}

\usepackage{hyperref}

\newcommand{\headrulewidth}{0pt}

\DeclareMathOperator{\E}{\mathbb E}

\newcommand{\bs}{\boldsymbol}

\usepackage{amsmath}
\usepackage{bbm}
\usepackage{amssymb}
\usepackage{color}
\usepackage{hyperref}
\usepackage[capitalize]{cleveref}
\usepackage{xr}
\usepackage{graphicx}

\usepackage{rotating}

\usepackage{ulem}
\usepackage{amsthm}
\usepackage{mathtools}
\usepackage{thmtools}
\usepackage{enumitem}
\usepackage[round]{natbib}
\usepackage{mathrsfs}
\usepackage{graphicx}
\usepackage{subcaption}
\usepackage{float}
\usepackage{booktabs,amsfonts,dcolumn}
\newcolumntype{d}[1]{D..{#1}}
\usepackage[T1]{fontenc}
\usepackage{parskip}
\usepackage[tracking=true]{microtype}
\usepackage{titlesec}

\newcommand{\nc}{\newcommand}
\nc{\red}{\color{red}}
\nc{\green}{\color{green}}
\nc{\vio}{\color{violet}}
\renewcommand{\qedsymbol}{$\blacksquare$}

\newcommand{\Fhat}{\hat F_{1,n}^{\text{ic},m}}
\newcommand{\Fstar}{\hat F_{1,n}^{*,\text{ic},m}}
\newcommand{\FONEmisig}{\mathcal{F}_1^\text{mi}}

\declaretheoremstyle[bodyfont=\slshape]{slshape}
\declaretheorem[style=slshape,name=Theorem,numberwithin=section]{thm}

\declaretheorem[style=slshape,name=Assumption,numberlike=thm]{ass}

\declaretheorem[style=slshape,name=Scheme,numberlike=thm]{scheme}

\newlist{thmlist}{enumerate}{1}
\setlist[thmlist]{label=(\roman{thmlisti}),noitemsep}

\Crefname{thm}{Theorem}{Theorems}
\Crefname{lem}{Lemma}{Lemmas}
\Crefname{ass}{Assumption}{Assumptions}
\Crefname{defi}{Definition}{Definitions}
\Crefname{rem}{Remark}{Remarks}
\Crefname{ex}{Example}{Examples}
\Crefname{cor}{Corollary}{Corollary}
\Crefname{repl}{Replacement}{Replacement}
\Crefname{scheme}{Scheme}{Scheme}

\Crefname{listthm}{Theorem}{Theorems}
\Crefname{listlem}{Lemma}{Lemmas}
\Crefname{listass}{Assumption}{Assumptions}
\Crefname{listdefi}{Definition}{Definitions}
\Crefname{listrem}{Remark}{Remarks}
\Crefname{listex}{Example}{Examples}
\Crefname{listcor}{Corollary}{Corollary}
\Crefname{listrepl}{Replacement}{Replacement}
\Crefname{listscheme}{Scheme}{Scheme}

\addtotheorempostheadhook[thm]{\crefalias{thmlisti}{listthm}}
\addtotheorempostheadhook[lem]{\crefalias{thmlisti}{listlem}}
\addtotheorempostheadhook[ass]{\crefalias{thmlisti}{listass}}
\addtotheorempostheadhook[defi]{\crefalias{thmlisti}{listdefi}}
\addtotheorempostheadhook[rem]{\crefalias{thmlisti}{listrem}}
\addtotheorempostheadhook[ex]{\crefalias{thmlisti}{listex}}
\addtotheorempostheadhook[cor]{\crefalias{thmlisti}{listcor}}
\addtotheorempostheadhook[repl]{\crefalias{thmlisti}{listrepl}}
\addtotheorempostheadhook[scheme]{\crefalias{thmlisti}{listscheme}}

\creflabelformat{listthm}{#2\thethm.#1#3}
\creflabelformat{listlem}{#2\thethm.#1#3}
\creflabelformat{listass}{#2\thethm.#1#3}
\creflabelformat{listdefi}{#2\thethm.#1#3}
\creflabelformat{listrem}{#2\thethm.#1#3}
\creflabelformat{listex}{#2\thethm.#1#3}
\creflabelformat{listcor}{#2\thethm.#1#3}
\creflabelformat{listrepl}{#2\thethm.#1#3}
\creflabelformat{listscheme}{#2\thethm.#1#3}

\allowdisplaybreaks
\title{{Inference via Wild Bootstrap and Multiple Imputation under Fine-Gray
Models with Incomplete Data }}

\author{Marina T.\ Dietrich\footnote{Department of Mathematics\\\phantom{X}\quad Faculty of Exact Science\\\phantom{X}\quad Vrije Universiteit Amsterdam\\\phantom{X}\quad De Boelelaan 1111\\\phantom{X}\quad 1081 HV Amsterdam\\\phantom{X}\quad The Netherlands} \!  \ Dennis Dobler$^*$, \ Mathisca C.\ M.\ de Gunst$^*$}

\date{\today}

\begin{document}
%\includepdf[pagecommand={\begin{tikzpicture}[remember picture, overlay];\end{tikzpicture}}]{BirgitSollie-Cover-Gekozen-BackRugCover}

\maketitle

\begin{abstract}
\noindent Fine-Gray models specify the subdistribution hazards for one out of multiple competing risks to be proportional.
The estimators of parameters and cumulative incidence functions under Fine-Gray models have a simpler structure when data are censoring-complete than when they are more generally incomplete.
This paper considers the case of incomplete data but it exploits the above-mentioned simpler estimator structure for which there exists a wild bootstrap approach for inferential purposes.
The present idea is to link the methodology under censoring-completeness with the more general right-censoring regime with the help of multiple imputation.
In a simulation study, this approach is compared to the estimation procedure proposed in the original paper by Fine and Gray when it is combined with a bootstrap approach.
An application to a data set about hospital-acquired infections illustrates the method.
\end{abstract}


%\newpage

\setstretch{1.3}
\newgeometry{left=3cm, right=2cm,bottom=3.5cm}

%\newpage
%\include{Acknowledgement/acknowledgement}

\renewcommand{\headrulewidth}{0.5pt}

%promotor:           prof.dr. M.C.M. de Gunst

%promotor:        prof.dr. M.R.H. Mandjes

%%%%%
%\newpage
%\vspace{100mm}
%\begin{center}
%Give thanks to the Lord, for he is good; \\
%his love endures forever. \\
%Psalm 107:1, NIV
%\end{center}
%%%%%

%\newpage\null\thispagestyle{empty}

% set the number of sectioning levels that get number and appear in the contents
%\setcounter{secnumdepth}{3}
%\setcounter{tocdepth}{1}

%\include{Dedication/dedication}
%\include{Abstract/abstract}

%\fancyfoot[C]{\thepage}

%\fancyhead[LE,RO]{\thepage}

%\tableofcontents
%\thispagestyle{empty}

%\printglossary[title=List of Acronyms,type=\acronymtype]
%\printglossary  % Print the nomenclature
%\addcontentsline{toc}{chapter}{Nomenclature}

%
%%
%%%
%%%%
%%%%%
%%%%%%
%%%%%%%
%\chapter{Inference via Wild Bootstrap and Multiple Imputation under Fine-Gray Models with Incomplete Data}
%\label{chap:MI}
%%%%%%%
%%%%%%
%%%%%
%%%%
%%%
%%
%
%
%%
%%%
%%%%
%%%%%
%%%%%%
%%%%%%%
\section{Introduction}
%%%%%%%
%%%%%%
%%%%%
%%%%
%%%
%%
%
The Fine-Gray model has been introduced in \cite{Fine-Gray} for semiparametric time-to-event analyses in competing risks settings. Moreover, the proposed methods depend on the available data: either complete data, censoring-complete data or incomplete data. The term complete data refers to the situation in which time-to-event data are collected in the absence of censoring. If censoring is present, the data are either called censoring-complete or incomplete. The difference between censoring-complete data and incomplete data is that censoring-complete data contain the censoring times for all individuals, as would be the case if censoring occurs only due to administrative loss to follow-up, whereas for incomplete data only the minimum of the event time and the censoring time is available. In other words, incomplete data do not contain the hypothetical censoring times of those individuals who have experienced either of the competing events.

In \cite{dietrich23} (Part~II), we have shown that the wild bootstrap is applicable to the estimators of the Fine-Gray model if the data are censoring-complete. In the case of incomplete data, one may use multiple imputation (MI) as suggested in \cite{ruan_gray} in order to generate the missing censoring times for those individuals who experienced an event. In this way, one obtains an augmented data set that is censoring-complete. One exemplary way for the step of data augmentation is  the Kaplan-Meier method proposed in \cite{taylor_MI}. Based on the augmented data set, inference for the cumulative incidence function (CIF) can be performed via the wild bootstrap. In this paper, we propose a novel wild bootstrap confidence band for the cumulative incidence function which is adapted to incomplete data via multiple imputation. In \cite{boot_MI}, and references therein, bootstrap estimation in combination with multiple imputation is investigated as well, but there it is used for interval estimation of a scalar parameter.

The idea of multiple imputation dates back to the 1970s when Rubin was working on missing data in the context of survey nonresponse, see, e.g., \cite{rubin_78}. In brief, multiple imputation is a statistical method which handles missing data by augmenting the data set with one potential value for each missing value in multiple iterations. Each iteration provides an augmented data set based on which the complete data estimator is computed. These imputation-based estimators are then combined by taking their average which is referred to as the overall estimator. It is crucial that the inference takes the uncertainty about the imputed values and the different imputation-based estimators into account. For example, in the case of a confidence interval for a scalar parameter the variance estimator of the overall estimator includes the so-called within-imputation variance and the so-called between-imputation variance. The former is the average over the variance estimators each calculated based on one of the augmented data sets, and the latter refers to the variability between the different imputation-based estimators. For more details on multiple imputation we refer to the text book \cite{rubin_87} and the review paper \cite{MI_review_murray}; for a general reference on statistical analysis with missing data, see  \cite{book_missing_data}.

%Incomplete data can also be modelled as coarsened data. Thus, one can resort to the ignorability condition ``coarsening at random'' for Bayes and likelihood inference. The concept of coarsened data and the corresponding condition has been established in \cite{Coarsening_concept} and it was applied to biomedical examples in \cite{Heitjan_93}. The extension to the frequentist ignorability condition ``coarsened completely at random'' can be found in \cite{Heitjan_94}. Here, coarsening at random and coarsening completely at random are generalizations of the concepts ``missing at random'' and ``missing completely at random'', respectively. See  \cite{MI_coarsened_data} for a reference on multiple imputation in the context of coarsened data.

In \cite{Fine-Gray} the proposed method to deal with incomplete data is inverse-probability-of-censoring-weighting (IPCW). This approach has been adopted by the authors of \cite{fastcmprsk} and connected with the bootstrap in order to provide confidence bands for the cumulative incidence function. We use IPCW
%of inverse-probability-of-censoring-weighting with the bootstrap
as an alternative method and compare the resulting bootstrap-based IPCW (B-IPCW) confidence bands to our proposed wild bootstrap-based multiple imputation (WB-MI) confidence bands in a simulation study. Additionally, we illustrate the two methods by means of a real data example.

This paper is organized as follows. In \cref{CC_data} we introduce the cumulative incidence function for Fine-Gray models and recap the corresponding wild bootstrap confidence bands based on censoring-complete data  studied in \cite{dietrich23} (Part~II). In \cref{subsec:imputation_estimates} we first describe the algorithm underlying the imputation of the missing censoring times and then define our novel WB-MI confidence band for the cumulative incidence function. The B-IPCW confidence band for the cumulative incidence function is presented in \cref{subsec:incomplete_estimates}. The comparison of the various methods via a simulation study can be found in \cref{sec:simulationStudy_Chp3}, and the corresponding illustration by a real data example in \cref{sec:data_ex_C3}. The real data example is accompanied by a sensitivity analysis regarding the multiple imputation method. Some concluding remarks are presented in \cref{sec:conclusion_Chp3}.

\iffalse
{\vio  However, the effect of multiple imputation on the point estimate and the wild bootstrap-based band estimate of the cumulative incidence function for the event of interest remains unclear. We take this to formulate our main research question in this chapter:
\\[0.2cm]
{\centering ``What is the influence of multiple imputation \\
of missing censoring times on wild bootstrap-based confidence bands \\
for semiparametric cumulative incidence functions?''}
\\[0.2cm]}
{\green to the above statement: the research question has changed. We now focus on bands and on how to construct WB bands under incomplete data via multiple imputation. Say that we asses the proposed bands via a simulation study and that we demonstrate its usefulness by means of a real data example.}

{\green check paper on imputation by Ed et.al. \cite{Ed_imputation} and\\
``On the consistency of supervised learning with missing values'' of
Julie Josse.}

\fi

%
%%
%%%
%%%%
%%%%%
%%%%%%
%%%%%%%
\section{Inference for CIFs with Censoring-Complete Data}\label{CC_data}
%%%%%%%
%%%%%%
%%%%%\label{CC_data}
%%%%
%%%
%%
%
When analyzing time-to-event data in a competing risks setting, the focus typically lies on the analysis of one of the competing events. Let us assume that there are $K$ competing event types and the event of type 1 is the event of interest. In order to study the incidence of type 1 events, we consider the so-called cumulative incidence function of type 1 for an individual with time-constant covariate vector $\textbf Z$ which is denoted by $F_1(\cdot|\textbf Z)$ and is defined as
\[F_1(t|\textbf Z) = P(T\leq t,\epsilon = 1 | \textbf{Z}),\quad t\in [0,\tau ],\]
where $T$ denotes the event time, $\epsilon$ denotes the event type, and $\tau$ is the maximum follow-up time of the study at hand. We assume throughout this section that the censoring time $C$ is conditionally independent of $T$ and $\epsilon$, given $\textbf{Z}$. For the Fine-Gray model the cumulative incidence function of event type 1 is given by
\[F_1(t|\textbf Z) = 1-\exp\big\{-\exp(\textbf Z^\top\bs{\beta}_0)A_{1;0}(t)\big\},\quad t\in[0,\tau],\]
where $\bs\beta_0$ is the true regression coefficient and $A_{1;0}$ is the true cumulative baseline subdistribution hazard of event type 1 (\cite{Fine-Gray}).
The point estimator of $F_1(\cdot|\textbf Z)$ based on censoring-complete data of $n$ individuals is denoted by $\hat{F}^{\text{cc}}_{1,n}(\cdot|\textbf Z) $ and is defined by
\[\hat{F}^{\text{cc}}_{1,n}(t|\textbf Z) = 1-\exp\big\{-\exp(\textbf Z^\top\hat{\bs\beta}^{\text{cc}}_n)\hat{A}^{\text{cc}}_{1;0,n}(t,\hat{\bs\beta}^{\text{cc}}_n)\big\},\quad t\in[0,\tau]. \]
Here, $\hat{\bs\beta}^{\text{cc}}_n$ is the maximum partial likelihood estimator of the regression coefficient $\bs\beta_0$ and $\hat{ A}^{\text{cc}}_{1;0,n}(\cdot,\hat{\bs\beta}^{\text{cc}}_n)$ is the Breslow estimator of the cumulative baseline subdistribution hazard $A_{1;0}$ both of which are defined based on censoring-complete data
\citep{Fine-Gray}.
%. See Sections~\ref{subsec:Fine-Gray}~and~\ref{subsec:F-G-estimators} for more details on the Fine-Gray model and the definitions of the corresponding estimators.

%\sloppy
Furthermore, we consider the unweighted and untransformed time-simultaneous $(1-\alpha)$-wild bootstrap confidence band for $F_1(\cdot|\textbf Z)$ presented in \cite{dietrich23} (Part~II). This wild bootstrap confidence band is based on censoring-complete data and we denote it by $CB^{*,\text{cc}}_{1,n}(\cdot|\textbf Z)$:
\begin{equation}
\label{eq:ccband}
CB^{*,\text{cc}}_{1,n}(t| \textbf Z) = \hat{F}^{\text{cc}}_{1,n}(t|\textbf Z) \mp q^{*,\text{cc}}_{1-\alpha , n}/\sqrt{n}, \quad t\in [t_1,t_2],
\end{equation}
where
\begin{itemize}
%    \item $n$ indicates the sample size;
    \item the boundaries of $[t_1, t_2]\subset [0,\tau]$ correspond to the first and the last decile of the observed survival times of the event of interest in order to avoid poor approximation of the underlying distribution at time points that relate to the extremes of the event times, cf.\ \cite{lin97};
    \item ${q}_{1-\alpha ,n}^{*,\text{cc}}$ is the conditional $(1-\alpha)$-wild bootstrap-quantile of
    $$\sup_{t\in[t_1,t_2]}\lvert \sqrt{n} (\hat{F}^{*,\text{cc}}_{1,n}(t | \textbf{Z}) - \hat{F}^{\text{cc}}_{1,n}(t | \textbf{Z}) )\rvert$$ given the censoring-complete data set. Here,
    $$\hat{F}^{*,\text{cc}}_{1,n}(t | \textbf{Z}) = 1-\exp\big\{-\exp(\textbf Z^\top\hat{\bs\beta}^{*,\text{cc}}_n)\hat A^{*,\text{cc}}_{1;0,n}(t,\hat{\bs\beta}^{*,\text{cc}}_n)\big\}, \quad t\in[0,\tau],$$
    is the wild bootstrap counterpart of $\hat F^{\text{cc}}_{1,n}(\cdot|\textbf Z)$), where $\hat{\bs\beta}^{*,\text{cc}}_n$ and $\hat A^{*,\text{cc}}_{1;0,n}(\cdot,\hat{\bs\beta}^{*,\text{cc}}_n)$ are the wild bootstrap versions of $\hat{\bs\beta}^{\text{cc}}_n$ and $\hat A^{\text{cc}}_{1;0,n}(\cdot,\hat{\bs\beta}^{\text{cc}}_n)$, respectively. We refer to \cite{dietrich23} (Part~II) for the definitions of the corresponding estimators. %Note that for the construction of the wild bootstrap estimators we have used standard normal multipliers.
\end{itemize}
Note that $CB^{*,\text{cc}}_{1,n}(\cdot|\textbf Z)$ is asymptotically valid, which means that the region between the lower bound and the upper bound of the confidence band over the time interval $[t_1,t_2]$ covers the entire cumulative incidence function $F_1(t|\textbf Z)$, $t\in [t_1,t_2]$, asymptotically with probability $(1-\alpha)$. Here, the term asymptotic refers to the sample size $n$ increasing to infinity. For more information on the mathematical background of the wild bootstrap confidence bands we refer to \cite{dietrich23} (Part~II).

%
%%
%%%
%%%%
%%%%%
%%%%%%
%%%%%%%
\section{Inference for CIFs with Incomplete Data via MI}\label{subsec:imputation_estimates}
%%%%%%%
%%%%%%
%%%%%
%%%%
%%%
%%
%
As mentioned above, multiple imputation is commonly used for scalar-valued estimands.
In this section, we extend multiple imputation-based inference to  function-valued estimands via time-simultaneous confidence bands. In particular, we propose in \cref{sec:WB_MI} a novel wild bootstrap confidence band for the cumulative incidence function that is adapted to incomplete data using multiple imputation. For this we present in \cref{subsec:3imputations}  the imputation scheme according to which the missing censoring times will be created.
In~\cref{subsec:3asymptotic} we will discuss the  asymptotic properties of the confidence bands.
%
%%
%%%
%%%%
%%%%%
%%%%%%
%%%%%%%
\subsection{Imputation of the Missing Censoring Times}\label{subsec:3imputations}\quad \\
%%%%%%%
%%%%%%
%%%%%
%%%%
%%%
%%
%
Given an incomplete data set, we use multiple imputation to create several augmented data sets which additionally contain one imputed censoring time for each individual who has experienced an event. Thus, the resulting augmented data sets are censoring-complete. As before, we assume that the censoring time $C$ is conditionally independent of $T$ and $\epsilon,$ given $\textbf{Z}$.
We present three multiple imputation techniques, where all of them follow the same scheme and differ only in the way the censoring survival distribution is estimated. The general imputation scheme has been proposed for the imputation of missing event times in \cite{taylor_MI} and has been used for the imputation of missing censoring times in the context of the Fine-Gray model in \cite{ruan_gray}. It consists of the following steps.
\begin{scheme}\label{WB-MI_scheme}\quad
%\newline
%\phantom{X}\\
\begin{thmlist}
\item Estimate the survival distribution $G(t) = \mathbb{P}(C>t)$ of the censoring time $C$. The corresponding estimate is denoted by $\hat{G}(t)$, $t\in [0,{\tau}].$%, where the upper bound $\tilde{\tau}$ of the support of $\hat{G}$ is determined by the method by which $G(t)$ is estimated.
    \label{item:imputation_step1}
    \item Impute a potential censoring time for each individual $i$ who has experienced an event by drawing from the estimated survival function conditionally on  the information that censoring has happened after the observed event, i.e. $\hat{\mathbb{P}}(C>t|C>T_i) = \hat{G}(t)/\hat{G}(T_i)$ for $t>T_i$, where $T_i$ is the observable event time of individual $i$.\label{item:imputation_step2}
    \item Augment the incomplete data set with the imputed censoring times such that the resulting data set is censoring-complete.\label{item:imputation_step3}
\end{thmlist}
\end{scheme}
The three  methods that we will use to estimate the survival distribution $G(t)$ in step~\ref{item:imputation_step1} of \cref{WB-MI_scheme} are: estimate $G(t)$ by means of the Kaplan-Meier methodology, via a Cox model, or according to a suitable parametric distribution. First, in case of the Kaplan-Meier methodology, $\hat G$ equals the Kaplan-Meier estimator of the censoring survival distribution, see \cite{taylor_MI}. The second approach has been suggested by Ruan and Gray for the case in which the censoring distribution depends on covariates (cf.\ \cite{ruan_gray}). In that case, we obtain $\hat{G}(t|\textbf{Z}_i) = \hat{\mathbb{P}}(C>t|\textbf{Z}_i) $, $\textbf{Z}_i$ being the covariate vector of individual $i$, by fitting a Cox proportional hazards model to the censoring times. For the third imputation technique, we propose to use a parametric distribution to estimate $G(t)$, such as the Weibull distribution or the uniform distribution. For this, we recommend to choose the parameters of the corresponding distribution in accordance with the Kaplan-Meier estimator of the censoring survival distribution.
The parametric choice can be used to ensure imputations free of ties by using continuous parametric distributions.

Finally, we repeat steps \ref{item:imputation_step2} and  \ref{item:imputation_step3} of \cref{WB-MI_scheme} $M$ times in order to obtain $M$ augmented data sets that are censoring-complete.

\subsection{WB-MI Confidence Bands}\label{sec:WB_MI}
%%%%%%%
%%%%%%
%%%%%
%%%%
%%%
%%
%
Suppose that, given an incomplete data set, we have created $M$ augmented data sets according to \cref{WB-MI_scheme}. We denote the incomplete data set by $\mathcal{D}$, so that
$$\mathcal{D}=\{\min (T_i , C_i), \mathbbm{1}\{T_i \leq C_i \}, \mathbbm{1}\{T_i \leq C_i \}\epsilon_i, \textbf{Z}_i, i=1\ldots , n \},$$
where the index $i\in\{1,\ldots , n \}$ refers to the $i$-th individual.
The augmented data sets are given by $\mathcal{D}^{1},\ldots ,\mathcal{D}^{M} $, that is, for $m=1,\ldots,M$,
$$\mathcal{D}^{m}=\{\min (T_i , C_i), \mathbbm{1}\{T_i \leq C_i \}, \mathbbm{1}\{T_i \leq C_i \}\epsilon_i, \textbf{Z}_i, \mathbbm{1}\{T_i \leq C_i \}C_i^{\text{ic},m} ,  i=1\ldots , n \},$$
where $C_i^{\text{ic},m}$ is the $m$-th imputed censoring time for individual $i$ for whom an event has been observed, $i\in\{1,\ldots , n\}$, and the superscript $\text{ic}$ stands for imputed censoring. Based on an augmented data set $\mathcal{D}^{m}$, the $m$-th imputation-based maximum partial likelihood estimator and the $m$-th imputation-based Breslow estimator are defined as it was the case for censoring-complete data. We denote the corresponding estimators by $\hat{\bs\beta}_n^{\text{ic},m}$ and $\hat{A}^{\text{ic},m}_{1;0,n}(t,\hat{\bs\beta}_n^{\text{ic},m})$, respectively. Additionally, we denote the $m$-th imputation-based estimator of the cumulative incidence function by
$$\hat{F}_{1,n}^{\text{ic},m}(t|\textbf Z) = 1-\exp\big\{-\exp(\textbf Z^\top\hat{\bs\beta}^{ic,m}_n)\hat{A}_{1;0,n}^{ic,m}(t,\hat{\bs\beta}^{ic,m}_n)\big\}, \quad t\in [0,\tau], \quad m=1,\ldots , M. $$
We define the MI estimator of the cumulative incidence function by taking the average over all $M$ imputation-based estimators of the cumulative incidence function, so that the MI estimator with respect to all $M$ augmented data sets, denoted by $\overline{F}_{M}^{\text{mi}}(\cdot|\textbf Z) $, is given by
\begin{align}
\label{Fbar-miM}
    \overline{F}_{M}^{\text{mi}}(t|\textbf Z) = \frac{1}{M}\sum_{m=1}^{M}\hat{F}_{1,n}^{\text{ic},m}(t|\textbf Z), \quad t\in [0,\tau],
\end{align}
where the superscript mi indicates that the corresponding estimator has been obtained via multiple imputation. Note that in the literature this estimator is called the overall estimator. Later in this section, we will see that for the wild bootstrap we condition on all $M$ augmented data sets. That is, we will condition on the following $\sigma$-algebra:
\[\mathcal{F}_1^{\text{mi}} =\sigma \{\mathcal{D}^{1}, \ldots ,\mathcal{D}^{M} \} .\]
$\overline{F}_{M}^{\text{mi}}$ is $\mathcal{F}_1^{\text{mi}} $-measurable and the additional randomness induced by the multiple imputation thus disappears when conditioning on that $\sigma$-algebra. In order to preserve the randomness of the multiple imputation in the wild bootstrap counterpart of the MI estimator of the cumulative incidence function, we propose to create this estimator with respect to a random subset of the augmented data sets $\mathcal{D}^{1},\ldots ,\mathcal{D}^{M} $.
For this, we draw $I< M$ out of the $M$ augmented data sets with replacement. We denote the corresponding data sets by $\mathcal{D}^{\pi (1)},\ldots ,\mathcal{D}^{\pi (I)}$, where $\pi: \{1,\ldots , I\} \mapsto \{1,\ldots , M\}$ maps the $l$-th draw to the index of the chosen data set, $l=1,\ldots , I$. Moreover, we denote the resulting WB-MI estimator of the cumulative incidence function by $\overline{F}_{I}^{*,\text{mi}}$, where this estimator will be defined below. Since we consider the WB-MI estimator of the cumulative incidence function with respect to this random subset of augmented data sets, we will mirror this in case of the ``basic'' MI estimator of the cumulative incidence function. In this way we ensure that the (conditional) covariances of the WB-MI estimator of the cumulative incidence function and the MI estimator of the cumulative incidence function coincide. With this motivation in mind, we introduce the MI estimator of the cumulative incidence function with respect to $I$ out of the $M$ augmented data sets:
\begin{align}
\label{Fbar-miI}
\overline{F}_{I}^{\text{mi}}(t|\textbf Z) = \frac{1}{I}\sum_{l=1}^{I}\hat{F}_{1,n}^{\text{ic},\pi ( l)}(t|\textbf Z), \quad t\in [0,\tau] .
\end{align}
Next, we wish to define the WB-MI estimator of the cumulative incidence function $\overline{F}_{I}^{*,\text{mi}}$. For this purpose, we define the wild bootstrap counterpart of the $\pi (l)$-th imputation-based estimators as the corresponding censoring-complete estimators based on the augmented data set $\mathcal{D}^{\pi ( l)}$, $l=1,\ldots , I$. See \cite{dietrich23} (Part~II) for the definitions of the wild bootstrap estimators based on censoring-complete data. In particular, we denote the wild bootstrap counterparts of the $\pi (l)$-th imputation-based maximum partial likelihood estimator $\hat{\bs\beta}_n^{\text{ic},\pi ( l)}$ and of the $\pi (l)$-th imputation-based Breslow estimator $\hat{A}^{\text{ic},\pi ( l)}_{1;0,n}(t,\hat{\bs\beta}_n^{\text{ic},\pi (l)})$ by $\hat{\bs\beta}_n^{*,\text{ic},\pi (l)}$ and $\hat{A}^{*,\text{ic},\pi (l)}_{1;0,n}(t,\hat{\bs\beta}_n^{*,\text{ic},\pi (l)})$, respectively. Moreover, we define the wild bootstrap counterpart of the $\pi (l)$-th imputation-based estimator of the cumulative incidence function by
$$\hat{F}_{1,n}^{*,\text{ic},\pi ( l )}(t|\textbf Z) = 1-\exp\big\{-\exp(\textbf Z^\top\hat{\bs\beta}^{*,ic,\pi ( l )}_n)\hat{A}_{1;0,n}^{*,ic,\pi ( l )}(t,\hat{\bs\beta}^{*,ic,\pi ( l )}_n)\big\}, \quad t\in [0,\tau], $$
$ l = 1,\ldots ,I.$
Note that the wild bootstrap estimators $\hat{\bs\beta}_n^{*,\text{ic},\pi (l)}$ and $\hat{A}^{*,\text{ic},\pi (l)}_{1;0,n}(t,\hat{\bs\beta}_n^{*,\text{ic},\pi ( l )})$ are based on the same multipliers $G_1,\ldots , G_n$ for all $ l=1,\ldots , I$.
This reflects the situation for the basic estimators $\hat{\bs\beta}_n^{\text{ic},\pi (l)}$ and $\hat{A}^{\text{ic},\pi (l)}_{1;0,n}(t,\hat{\bs\beta}_n^{\text{ic},\pi (l)})$ which are computed based on the same data set $\mathcal{D}$ for all realized imputations.
%\newpage
Finally, the WB-MI estimator of the cumulative incidence function with respect to $I$ out of the $M$ augmented data sets is defined as
\begin{align}
\label{Fbar*miI}
    \overline{F}_{I}^{*,\text{mi}}(t|\textbf Z) = \frac{1}{I}\sum_{l=1}^{I}\hat{F}_{1,n}^{*,\text{ic},\pi (l)}(t|\textbf Z), \quad t\in [0,\tau].
\end{align}
%\newpage
We are now ready to present the novel time-simultaneous $(1-\alpha)$-wild bootstrap confidence band for $F_1(\cdot|\textbf Z)$ which is computable from incomplete data via multiple imputation. Analogously to the band \eqref{eq:ccband}, this confidence band is untransformed and unweighted, and we denote it by  $CB^{*,\text{mi}}_{1,n}$. This WB-MI  $(1-\alpha)$ confidence band is defined as
\begin{equation}
    \label{eq:miband}
CB^{*,\text{mi}}_{1,n}(t| \textbf Z) = \overline{F}^{\text{mi}}_{I}(t|\textbf Z) \mp q^{*,\text{mi}}_{1-\alpha , n}/\sqrt{n}, \quad t\in [t_1,t_2],
\end{equation}
where
${q}_{1-\alpha ,n}^{*,\text{mi}}$ is the conditional $(1-\alpha)$-wild bootstrap-quantile of
    $$\sup_{t\in[t_1,t_2]}\lvert \sqrt{n} (\overline{F}^{*,\text{mi}}_{I}(t | \textbf{Z}) - \overline{F}^{\text{mi}}_{M}(t | \textbf{Z}) )\rvert$$
given $\mathcal{F}_1^{\text{mi}}$.
%Similarly as in {\green Section 3 of chapter 2}, it is also possible to produce transformed and weighted confidence bands. In future research, we intend to provide these confidence bands as well.

%%%%%%%%%%%

\subsection{Asymptotic Considerations for WB-MI Confidence Bands}
\label{subsec:3asymptotic}
\noindent
We now consider the asymptotic validity of the proposed WB-MI confidence bands.
For the bands to be  asymptotically valid, one would need to show the asymptotic equality of the distribution of
\begin{equation}
\label{process1}\sqrt{n}( \overline{F}^{\text{mi}}_{I}(t|\textbf Z) - {F}_{1}(t|\textbf Z)), \quad t\in[0,\tau],
\end{equation}
and the conditional distribution, given $\mathcal{F}^{\text{mi}}_1 $, of
\begin{equation}
\label{process2}\sqrt{n}( \overline{F}^{*,\text{mi}}_{I}(t|\textbf Z) - \overline{F}^{\text{mi}}_{M}(t|\textbf Z)), \quad t\in[0,\tau],
\end{equation}
in probability, as $I=I(n), M=M(n), n \to \infty$.
This means, weak convergence of these sequences of stochastic processes and equality of the limit processes.
%In view of the fact that $\overline{F}^{*,\text{mi}}_{I}$ is the average of $I$ conditionally independent processes on  $[0,\tau]$, the following assumption seems appropriate. \\
In order to motivate the weak convergence of these processes, we resort to the conditional weak convergence result for an exchangeably weighted bootstrapped statistic provided in \cite{Dobler_Pauly_14}.
%Furthermore, the authors of \cite{Dobler_Pauly_14} proved the conditional weak convergence of an exchangeably weighted bootstrapped statistic.
The structure of the weighted bootstrapped statistic with chosen weights related to Efron's bootstrap is similar to the structure of $\overline{F}^{\text{mi}}_{I}$ and $\overline{F}^{*,\text{mi}}_{I}$, cf.\ \eqref{eq:Form_FI__F1} and \eqref{diffFbars}, respectively.
Note that the centering term involved in (3.3) of \cite{Dobler_Pauly_14} is missing in \eqref{process1} and \eqref{process2}.
A rectification of this in the present context will result in a change of covariances but correcting for the centering term usually does not pose any problems in terms of the tightness of the sequence of stochastic processes.
In view of the that motivation, the upcoming \cref{assumpGaussian} seems appropriate.
\begin{ass}\label{assumpGaussian}
For $I, M, n  \to \infty$, each of the processes
$\sqrt{n}( \overline{F}^{\text{mi}}_{I}(t|\textbf Z) - {F}_{1}(t|\textbf Z))$
and, conditionally on $\mathcal{F}_1^{\text{mi}}$,
$\sqrt{n}( \overline{F}^{*,\text{mi}}_{I}(t|\textbf Z) - \overline{F}^{\text{mi}}_{M}(t|\textbf Z))$,  $t\in [0,\tau]$, weakly converges to some zero-mean Gaussian process. The latter convergence is understood to hold in probability.
\end{ass}
%\ \\
%\smallskip
%\noindent
Under \cref{assumpGaussian}, for proving equality of the asymptotic distributions of the two processes, it suffices to show  the equality of their (conditional) expectations and (conditional) (co-)variances. For this we make the following set of assumptions.
%\newpage
\begin{ass}
\label{assumptions_C3}
\begin{align}
 \label{ass:mom0} & \E(\hat F_{1,n}^{\text{ic},1})  = F_1 + o(n^{-1}), \\
 \label{ass:mom0*} & \E(\Fstar| \FONEmisig)  = \Fhat + o_p(n^{-1})
 %\\
 % & \nonumber \qquad\qquad
% \qquad
  \text{\ with } \E(| o_p(n^{-1}) | ) = o(n^{-1}), \\
  \nonumber & \qquad\qquad\qquad\qquad\qquad\qquad\qquad \text{uniformly in } m =1,\dots, M,\\
 \label{ass:mom1} & \text{Var}(\hat F_{1,n}^{\text{ic},1})  = \frac{V}{n} + o(n^{-1}), \\
 %\label{ass:mom1*} & \text{Var}(\Fstar | \FONEmisig)  = \text{Var}(\hat F_{1,n}^{\text{ic},m}) + o_p(n^{-1})
 %\\
 %\nonumber & \qquad\qquad
 %\ \  \text{uniformly in } m =1,\dots, M, \\
 \label{ass:mom21} & \text{Cov}(\hat F_{1,n}^{\text{ic},1}, \hat F_{1,n}^{\text{ic},2})  = \frac{C}{n} + o(n^{-1}), \\
 %\label{ass:mom1*} \E((\hat F_{1,n}^{*,\text{ic},m})^2 | \FONEmisig) & = (\hat F_{1,n}^{\text{ic},m})^2 + O_p(n^{-1/2}) \quad \text{uniformly in } m =1,\dots, M, \\
 \label{ass:mom21*} & \text{Cov}(\hat F_{1,n}^{*,\text{ic},m_1}, \hat F_{1,n}^{*,\text{ic},m_2} | \FONEmisig)  = \text{Cov}(\hat F_{1,n}^{\text{ic},m_1}, \hat F_{1,n}^{\text{ic},m_2}) + o_p(n^{-1}) \\
 \nonumber & \qquad\qquad\qquad\qquad\qquad\qquad\qquad \text{uniformly in } m_1,m_2 =1,\dots, M, \\
 %\label{ass:mom_eps} & \sum_{m=1}^M \big( \E( \Fstar | \FONEmisig ) - \E(\hat F_{1,n}^{\text{ic},1}) \big)  = O_p(M^{1-\varepsilon}) \quad  \text{ for some } 0<\varepsilon<1. \\
 \label{ass:mom_eps2} &  \sum_{m=1}^M \big(  \Fhat  - F_1 \big)  = O_p(M^{1/2}) + O_p(M n^{-1/2}) . \\
 \label{ass:mom_eps3} & \sum_{m=1}^M \big(  (\Fhat)^2  - F_1^2 \big)  = O_p(M^{1/2})  + O_p(M n^{-1/2}) .
\end{align}
\end{ass}
\noindent
Here, $V$ and $C$ are variance and covariance functions defined on $[0,\tau]$ and $[0,\tau]^2$, respectively, with $V< \infty$ and $C< \infty$ on the corresponding intervals.
%\noindent
%Without loss of generality, we assume for $\varepsilon$ in~\eqref{ass:mom_eps} that $\varepsilon \in (0,\tfrac12)$.
%Without loss of generality, we assume the $\varepsilon$'s in \eqref{ass:mom_eps}, \eqref{ass:mom_eps2}, and \eqref{ass:mom_eps3} to be the same by picking the smallest of the three.
In addition, we assume that $I$ and $M$ with $I \leq M$ are chosen depending on the sample size in such way that, as $n, I=I(n), M=M(n) \to \infty$,
\begin{align}
 \label{ass:I} \frac{n}{I^2} & = o(1), \\ %\to 0, \\
 \label{ass:M} \frac{I^2}{M} & = O(1). %\to 0, \quad \text{for $\varepsilon$ from \eqref{ass:mom_eps}}.
\end{align}
The assumptions \eqref{ass:mom0} and \eqref{ass:mom0*} are plausible in the light of
%the assumed (conditional) weak convergences for the (wild %bootstrap) incomplete estimators $\hat %F_{1,n}^{\text{ic}} $ and $\hat F_{1,n}^{*,\text{ic}} $,
\cref{assumpGaussian}.
%and because the imputation is random.
% DD: Why is the randomness of the imputation important here??
Assumptions \eqref{ass:mom1},  \eqref{ass:mom21}, and \eqref{ass:mom21*} are common in the context of central limit theorems.
For the  stochastic bound~\eqref{ass:mom_eps2} we remark that
conditionally on $\mathcal{F}_1$, the sum contains independent terms with the same conditional expectation $\E(\hat F_{1,n}^{\text{ic,1}} | \mathcal{F}_1 ) $, where $\mathcal{F}_1 = \sigma \{\mathcal{D}\}$.
As a consequence, it seems plausible to assume that
as $M\to \infty$ a conditional central limit theorem for
$$ \frac{1}{M} \sum_{m=1}^M \big( \Fhat - \E(\hat F_{1,n}^{\text{ic,1}} | \mathcal{F}_1 )) $$
holds in probability. Hence, the assumed stochastic boundedness by $O_p(M^{1/2})$.
This statement then clearly also holds unconditionally.
Furthermore, the reasoning behind the convergence rate of the remainder term  $M \cdot (\E(\hat F_{1,n}^{\text{ic},1} | \mathcal{F}_1 ) - F_1)$ is as follows.
Suppose that for $M$ large enough $\E(\overline{F}_{M}^{\text{mi}} | \mathcal{F}_1 )$ is approximately equal to a $\sqrt{n}$-consistent estimator $\tilde{F}_{1,n}$ for $F_1$.
One instance of such a $\mathcal{F}_1$-measurable estimator is the IPCW estimator $\hat{F}^{\text{ipcw}}_{1,n}$ that will be introduced in \cref{CB_boot_IPCW}. See Appendix B of \cite{Fine-Gray} for its convergence rate.
%Furthermore, suppose that $\E(\overline{F}_{M}^{\text{mi}} | \mathcal{F}_1 ) \approx \hat F_{1,n}^{\text{ipcw}}$ for $M$ large enough.
As a motivation for $\E(\overline{F}_{M}^{\text{mi}} | \mathcal{F}_1 ) \approx \tilde{F}_{1,n}$ for $M$ large enough, we refer to (3.5) of \cite{MI_review_murray}.
Combining this approximation with the fact that $\hat{F}^{\text{ic},1}_{1,n},\ldots ,\hat{F}^{\text{ic},M}_{1,n}$ are identically distributed, yields $\E({F}_{1,n}^{\text{ic},1} | \mathcal{F}_1 ) \approx\tilde{F}_{1,n}$.
Thus, we approximate the remainder term  $M \cdot (\E(\hat F_{1,n}^{\text{ic},1} | \mathcal{F}_1 ) - F_1)$ by $M \cdot (\tilde{F}_{1,n} - F_1)$ for which the convergence rate equals $M n^{-1/2}$.
%For the remainder term,  $M \cdot (\E(\hat F_{1,n}^{\text{ic},1} | \mathcal{F}_1 ) - F_1)$, we may assume a similar convergence rate as before, hence the rate $M n^{-1/2}$.
For \eqref{ass:mom_eps3} the reasoning is analogous.
Note that it might depend on the quality of the imputation method whether the assumptions above are satisfied.

%is sufficient to specify the corresponding Gaussian limit distributions.
%%
%%Subsequently, we assume the weak convergence towards some Gaussian processes.
%%In this way, our heuristic motivation consists of showing the equality of the (conditional) expectations and (conditional) variances under some additional regularity assumptions.
%%
%Thus, we will show under some additional regularity assumptions the equality of the (conditional) expectations and (conditional) variances.
%it is left to show the equality of the (conditional) expectations and (conditional) (co-)variances.
%in order to specify the limiting distribution completely.
%Under some additional regularity assumptions, we will provide the corresponding arguments for the (conditional) expectations and (conditional) variances.
%we will analyze the (conditional) expectations and (conditional) variances.
With the above assumptions, the following asymptotic equivalence result for the (conditional) expectations and (conditional) (co)variances of \eqref{process1} and \eqref{process2} can be proved.
For reasons of similarity, we will only consider the expectations and variances and omit the covariances.
%Under this assumption, it is sufficient to analyze their (conditional) expectations and (conditional) (co-)variances in order to completely specify the corresponding Gaussian processes. For reasons of similarity, we will only focus on the asymptotic variances and not on the covariances.
%Next, under some regularity assumptions, we will prove the asymptotic equality of the (conditional) variances (given $\mathcal{F}_1$) to argue the asymptotic validity of the proposed approach.
%%
%%
%%
\iffalse
\begin{ass}\label{assumptions_C3}\\
\begin{thmlist}
\item $\mathbb{E}\big(\sqrt{n}(\hat{F}^{*,\text{ic},m}_{1,n}(t|\textbf Z) - \hat{F}^{\text{ic},m}_{1,n}(t|\textbf Z)) |\mathcal{F}_1^{\text{mi}} \big)\stackrel{\mathbb{P}}{\longrightarrow}0, \text{ as } n\rightarrow \infty$;
\item $\mathbb{E}\big (  \sqrt{n} (\hat{F}^{\text{ic},1}_{1,n}(t|\textbf Z) - F_1(t|\textbf Z))\big )\longrightarrow 0, \text{ as } n\rightarrow \infty$;
\item $n\text{Var}\big(\hat{F}^{*,\text{ic},m}_{1,n}(t|\textbf Z) |\mathcal{F}_1^{\text{mi}} \big)\stackrel{\mathbb{P}}{\longrightarrow} \Sigma(t), \text{ as } n\rightarrow \infty$;
\item $n\text{Var}\big (  \hat{F}^{\text{ic},1}_{1,n}(t|\textbf Z) \big )\longrightarrow  \Sigma(t), \text{ as } n\rightarrow \infty$;
\end{thmlist}
where $\Sigma(t)$ is some covariance function.
\end{ass}
\fi
%%
%%
%%
\begin{thm}
\label{thm:asymptotic_equivalence}
If Assumptions \eqref{assumpGaussian} and \eqref{assumptions_C3}  hold, then we have for each fixed $t\in[0,\tau]$,
\begin{align}\label{eq:statement_E}
    \mathbb{E}\big( \sqrt{n}( \overline{F}^{*,\text{mi}}_{I}(t|\textbf{Z}) - \overline{F}^{\text{mi}}_{M}(t|\textbf{Z}))|\mathcal{F}_1^{\text{mi}}\big) - \mathbb{E}\big(\sqrt{n}( \overline{F}^{\text{mi}}_{I}(t|\textbf{Z}) - {F}_{1}(t|\textbf{Z})) \big) \stackrel{\mathbb{P}}{\longrightarrow} 0
\end{align}
and
\begin{align}\label{eq:statement_Var}
    \text{Var}\big( \sqrt{n}( \overline{F}^{*,\text{mi}}_{I}(t|\textbf{Z}) - \overline{F}^{\text{mi}}_{M}(t|\textbf{Z}))|\mathcal{F}_1^{\text{mi}}\big) - \text{Var}\big(\sqrt{n}( \overline{F}^{\text{mi}}_{I}(t|\textbf{Z}) - {F}_{1}(t|\textbf{Z})) \big) \stackrel{\mathbb{P}}{\longrightarrow} 0,
\end{align}
as $I, M, n\rightarrow \infty$ .
\begin{proof}
See Appendix.
\end{proof}
\end{thm}
\section{Inference for CIFs with Incomplete Data via IPCW}\label{subsec:incomplete_estimates}%\quad
%\\
%%%%%%%
%%%%%%
%%%%%
%%%%
%%%
%%
%

In this section, we present an alternative, existing, approach to inference for the cumulative incidence function with incomplete data. In particular, we consider an estimator of the cumulative incidence function for incomplete data which is obtained via IPCW and for which the corresponding inference is based on bootstrapping.

\subsection{IPCW in Fine-Gray Models}

For incomplete data the authors of \cite{Fine-Gray} adjusted the estimators involved in their model by means of the IPCW technique, cf.\ \cite{IPCW}. Here, a crucial role is played by a weight function due to which the unobserved censoring times are taken into account. Recall that for censoring-complete data the censoring times are available for those individuals who have experienced a competing event and that their at-risk indicator switches from one to zero at the time of censoring. In case of incomplete data, the censoring times are not available for these individuals and the corresponding at-risk indicators are constructed by IPCW as follows. The at-risk indicator based on complete data, denoted by $\tilde{Y}_i(t)$, is multiplied by a weight function, denoted by $w_i(t)$, such that the weighted at-risk indicator $w_i(t) \tilde{Y}_i(t)$ of an individual $i$ who experienced a competing event equals one until the occurrence of the competing event ($t\leq T_i$), and it is equal to the Kaplan-Meier estimator of ${\mathbb{P}}(C>t|C>T_i)$ after the occurrence of the competing event ($t>T_i$).
%Here, $C$ is the censoring time and $T_i$ is the event time of individual $i$ who has experienced the competing event.
In this manner, the weighted at-risk indicator decreases with the conditional survival distribution of the censoring time and the censoring time is accounted for. Furthermore, the values of the weighted at-risk indicators for those individuals who either experienced the event of interest or who have been censored, are equal to the values of the at-risk indicators based on censoring-complete data.
Additionally, the values of the weighted complete-data counting process increments of any individual are identical to the values of the corresponding counting-process increments for censoring-complete data.
%We refer to \cref{subsec:F-G-estimators} for the counting-process notation and at-risk indicator for censoring-complete data and to \cref{CB_boot_IPCW} below for the definition of the corresponding quantities for complete data. Furthermore, we assume in this section that the censoring time $C$ is independent of $T,\epsilon,$ and $\textbf{Z}$.
In this section, we assumed that the censoring time $C$ is independent of $T,\epsilon,$ and $\textbf{Z}$.

\subsection{B-IPCW Confidence Bands}\label{CB_boot_IPCW}
In the following, we present the IPCW estimator of the cumulative incidence function for incomplete data as defined in \cite{Fine-Gray}:
\[\hat{F}^{\text{ipcw}}_{1,n}(t|\textbf Z) = 1-\exp\big\{-\exp(\textbf Z^\top\hat{\bs\beta}^{\text{ipcw}}_n)\hat{A}^{\text{ipcw}}_{1;0,n}(t,\hat{\bs\beta}^{\text{ipcw}}_n)\big\},\quad t\in[0,\tau]. \]
Here, $\hat{\bs\beta}^{\text{ipcw}}_n$ is the maximum pseudo likelihood estimator for the regression coefficient $\bs\beta_0$, and $\hat A_{1;0,n}^{\text{ipcw}}(\cdot,\hat{\bs\beta}^{\text{ipcw}}_n)$ is a variant of the Breslow estimator for the cumulative baseline subdistribution hazard $A_{1;0}$. In particular, $\hat{\bs\beta}^{\text{ipcw}}_n$ is the root of the score statistic $\tilde{\textbf U}_n(t,\bs\beta) $ evaluated at $t=\tau$ with
\[{\tilde{\textbf U}}_n(t,\bs\beta) = \sum_{i=1}^n \int_0^t \big(\textbf Z_i - \frac{\sum_j w_j(u)\tilde{Y}_j(u)\textbf Z_j\exp \{ \textbf Z_j^\top{\bs\beta}\}}{\sum_j w_j(u)\tilde{Y}_j(u)\exp \{ \textbf Z_j^\top{\bs\beta}\}}\big) w_i(u)d\tilde{N}_i(u), \quad t\in[0,\tau].\]
Here, %$\textbf Z_i$ is the covariate vector of individual $i$,
$\tilde{N}_i(t)=\mathbbm{1}\{T_i \leq t, \epsilon_i = 1\} $ is the counting process for complete data, $\tilde{Y}_i(t)= 1~-~\tilde{N}_i(t-)$ is the at-risk indicator for complete data, and $w_i(t) = r_i(t)\hat{G}(t)/\hat{G}(\min(T_i,C_i,t))$ is the weight function with so-called vitality status $r_i(t) = \mathbbm{1}\{C_i\geq \min (T_i,t)\}$ and Kaplan-Meier estimator $\hat{G}$ of $G(t) = \mathbb{P}(C\geq t)$.
Note that for individuals with vitality status equal to one, it holds that
\[w_i(t) = \hat{G}(t)/\hat{G}(\min(T_i,C_i,t)) =
\begin{cases}
        1,\text{ if } t\leq T_i, \\
        \hat{\mathbb{P}}(C>t|C>T_i), \text{ if } t> T_i  .
        \end{cases}
\]
Additionally, the variant of the Breslow estimator is given by
\[\hat A^{\text{ipcw}}_{1;0,n}(t,\hat{\bs\beta}^{\text{ipcw}}_n) = \frac{1}{n}\sum_{i=1}^n \int_0^t \big( \frac{1}{n}\sum_{i=1}^n w_i(u)\tilde{Y}_i(u) \exp \{ \textbf Z_i^\top\hat{\bs\beta}^{\text{ipcw}}_n\}  \big)^{-1} w_i(u)d\tilde{N}_i(u),  \quad t\in[0,\tau].\]
%Note that $w_i(t)d\tilde{N}_i(t) = 1$ for $t=T_i$ if and only if the event of interest is observed for individual $i$, i.e.\ $T_i < C_i$ and the event type is 1.

Furthermore, we consider an unweighted and untransformed time-simultaneous $(1-\alpha)$-bootstrap confidence band for $F_1(\cdot|\textbf Z)$ with respect to the same time interval $[t_1,t_2]$ as considered in \cref{CC_data}. We denote this B-IPCW confidence band by $CB^{\text{boot, ipcw}}_{1,n}(\cdot|\textbf Z)$. Analogously to the weighted and transformed B-IPCW confidence band for $F_1(\cdot|\textbf Z)$ presented in \cite{fastcmprsk}, $CB^{\text{boot, ipcw}}_{1,n}(\cdot|\textbf Z)$ is given by
\[CB^{\text{boot, icpw}}_{1,n}(t|\textbf Z) = \hat{F}^{\text{icpw}}_{1,n}(t | \textbf Z) \mp q^{\text{boot, icpw}}_{1-\alpha,n} / \sqrt{n},\quad t\in [t_1,t_2],\]
where $q^{\text{boot, icpw}}_{1-\alpha,n}$ is the conditional $(1-\alpha)$-bootstrap-quantile of
$$\sup_{t\in[t_1,t_2]} \lvert \sqrt{n}(\hat{F}^{\text{boot, ipcw}}_{1,n}(t | \textbf Z) - \hat{F}^{\text{ipcw}}_{1,n}(t | \textbf Z)) \rvert, $$
given the incomplete data set $\mathcal{D}$. Here, $\hat{F}^{\text{boot, ipcw}}_{1,n}(\cdot | \textbf Z)$ is the B-IPCW estimator of the cumulative incidence function which is defined analogously to $\hat{F}^{\text{ipcw}}_{1,n}(\cdot | \textbf Z)$ but based on a bootstrap sample $\mathcal{D}^{\text{boot}}$ of the original data set $\mathcal{D}$.

%
%%
%%%
%%%%
%%%%%
%%%%%%
%%%%%%%
\section{Simulation Study}
\label{sec:simulationStudy_Chp3}
%%%%%%%
%%%%%%
%%%%%
%%%%
%%%
%%
%
%
%%
%%%
%%%%
%%%%%
%%%%%%
%%%%%%%
\subsection{Simulation Set-Up}\label{sec:set-up_Chp3}
%%%%%%%
%%%%%%
%%%%%
%%%%
%%%
%%
%

The simulation study presented in this section is inspired by the \textit{sir.adm} data set of the \texttt{mvna} R-package and is performed using R-4.1.0, cf.\ \cite{R}. The goal is twofold: first, to assess the coverage probability of the proposed WB-MI $95\%$ confidence bands $CB^{*,\text{mi}}_{1,n}(\cdot | \textbf Z = \textbf{z}_0)$ for $F_1(\cdot | \textbf Z = \textbf{z}_0)$ with respect to various settings for some chosen covariate vector $\textbf{z}_0$.
Second, we wish to compare the corresponding coverage probabilities and widths with those of the B-IPCW confidence bands and with those of the benchmark wild bootstrap confidence bands based on censoring-complete data.
To this end, we compute the empirical coverage probabilities and the median widths across all simulation studies of a kind.
For the computation of the coverage probabilities of the wild bootstrap confidence bands for censoring-complete data we have simulated 5,000 studies for each simulation setting. For the two confidence bands based on incomplete data we have simulated 1,000 studies for each setting. Additionally, we have determined the quantiles for the wild bootstrap confidence bands with censoring-complete data based on 2,000 wild bootstrap iterations and for the two confidence bands based on incomplete data we have used 1,000 (wild) bootstrap iterations.
We have chosen these iteration numbers to obtain very precise results for the benchmark method (the censoring-complete estimator) and to make a trade-off between computational costs and simulation accuracy for the multiple imputation-based estimator.

The simulation settings underlying this simulation study were chosen as follows:
\begin{itemize}
    \item sample sizes: $n = 100, 200$;
    \item nominal confidence level: $1-\alpha = 95\%$;
    \item multiplier distribution: $\mathcal{N}(0,1)$, since this distribution in combination with the quantile ${q}_{1-\alpha,n}^{*,\text{cc}}$ led to the best results in the simulation study of \cite{dietrich23} (Part~II);
    \item censoring distributions: $\mathcal{U}(0,c)$  with varying maximum parameter $c$ such that the average censoring rate over all samples is either between $20\%$ and $25\%$ (light censoring) or between $37\%$ to $43\%$ (strong censoring);
    \item covariates: trivariate $(Z_{ij})_{j=1}^3$ with independent $Z_{i1} \sim \mathcal{N}(0,1)$, $Z_{i2} \sim \textnormal{Bernoulli}(0.15)$, $Z_{i3} \sim \textnormal{Bernoulli}(0.4)$, where $Z_{ij}$ corresponds to the standardized age ($j=1$), to the pneumonia status ($j=2$), and to the gender ($j=3$) of individual $i$, $i=1,\ldots ,n$;
    \item time-constant \textit{cause-specific} baseline hazard rates of event type 1 and of event type 2: $(\alpha^1_{01;0}, \alpha^1_{02;0}) = (0.08, 0.008)$ and $(\alpha^2_{01;0}, \alpha^2_{02;0}) = (0.05, 0.05)$. The first pair of rates is motivated from the \textit{sir.adm} data set, and the second pair was chosen with the intention to produce more type-2 events;
    \item parameter vector: $\bs{\beta}_0 = (-0.05,-0.25,-0.05)$;
    \item covariate choices for the confidence bands: $\textbf{z}_0 = (-2/3, 0, 1)$, namely, a 45 years old female without pneumonia.
    \item number of multiple imputations: $I=10$ and $M=10000$;
    \item method of imputation: Kaplan-Meier method using the \texttt{kmi} R-package \citep{kmi};
    \item B-IPCW confidence bands: computed using the \texttt{fastcmprsk} R-package \citep{fastcmprsk_R}.
\end{itemize}
The simulation of the survival times and event types according to the Fine-Gray model has been conducted as described in \cite{dietrich23} (Part~II). In particular, we used the algorithms proposed in \cite{Beyersmann} under the parameter choices stated above. Furthermore, the confidence bands are constructed with respect to a subset $[t_1,t_2]$ of $[0,\tau]$ in order to avoid poor approximation of the underlying distribution at time points that relate to the extremes of the event times, cf.\ \cite{lin97}.
Here, we have used for $t_1$ and $t_2$ the first and the last decile, respectively, of the observed survival times of event type 1 across all samples of a kind. For each sample, $t_1$ was also taken to be at least the first observed survival time of type 1.

Below, we will also take into account how the different censoring distributions and hazard rates affect the number of individuals for whom a missing censoring time needs to be imputed.
Moreover, we will analyze how this will affect the accuracy of the multiple imputation-based confindence bands.
Under $(\alpha^1_{01;0}, \alpha^1_{02;0})$, the fraction of type-1 events is about five times higher than the fraction of type-2 events resulting in a relatively low number of individuals for whom a potential censoring time is imputed. Combining these hazard rates with strong censoring led to a simulation setting for which the pool of censoring times from which to impute is the largest.
Combining $(\alpha^1_{01;0}, \alpha^1_{02;0})$ with strong censoring results in the most favorable simulation setting, that is, the least number of censoring times need to be imputed from the largest set of observed censoring times among all the tested settings.
In contrast, under $(\alpha^2_{01;0}, \alpha^2_{02;0})$, the fraction of type-1 events and type-2 events is similar. As a consequence, the number of individuals for which the missing censoring time is imputed under $(\alpha^2_{01;0}, \alpha^2_{02;0})$ is considerably higher compared to $(\alpha^1_{01;0}, \alpha^1_{02;0})$.
 Additionally, combining $(\alpha^2_{01;0}, \alpha^2_{02;0})$ with light censoring results in the least favorable setting for multiple imputation, because the censoring times of a relatively high number of individuals is imputed while the number of individuals from whom to impute is relatively low. In \cref{subsec:results_simStudy_Chp3} we will see whether these circumstances have an effect on the coverage probability or on the width of the corresponding WB-MI confidence bands.

%
%%
%%%
%%%%
%%%%%
%%%%%%
%%%%%%%
\subsection{Results of the Simulation Study}\label{subsec:results_simStudy_Chp3}
%%%%%%%
%%%%%%
%%%%%
%%%%
%%%
%%
%

In \cref{tab:sim_results_C3} the results of the simulation study are presented. Here, we show the coverage probability (CP) in percentage and the width of the confidence bands computed for either censoring-complete data (CC) or for incomplete data. In case of censoring-complete data, we compute confidence bands using the wild bootstrap (WB). Otherwise, we distinguish between WB-MI confidence bands and B-IPCW confidence bands. Additionally, each row of the table corresponds to one of the simulation settings described in \cref{sec:set-up_Chp3}.

\begin{table}[ht]
\centering
\begin{tabular}{lllllllll}
  \hline
  \hline
&  &  &  & CP &  &  & WIDTH &  \\
 &  &  & CC & \multicolumn{2}{c}{Incomplete}  & CC & \multicolumn{2}{c}{Incomplete} \\
  n & censoring & ($\alpha_{01;0},\alpha_{02;0}$) & WB & WB-MI & B-IPCW & WB & WB-MI & B-IPCW \\
  100 & light & (0.08,0.008) & 95.4 & 94.9 & 93.3 & 0.426 & 0.426 & 0.408 \\
   &  & (0.05,0.05) & 94.6 & 93.6 & 91.8 & 0.414 & 0.412 & 0.390 \\
   & strong & (0.08,0.008) & 94.4 & 94.0 & 93.2 & 0.455 & 0.454 & 0.435 \\
   &  & (0.05,0.05) & 94.1 & 94.6 & 92.6 & 0.422 & 0.423 & 0.395 \\
  200 & light & (0.08,0.008) & 94.9 & 95.8 & 95.2 & 0.295 & 0.295 & 0.289 \\
   &  & (0.05,0.05) & 93.8 & 93.6 & 92.7 & 0.285 & 0.286 & 0.280 \\
   & strong & (0.08,0.008) & 93.4 & 94.2 & 93.7 & 0.316 & 0.317 & 0.311 \\
   &  & (0.05,0.05) & 93.5 & 93.4 & 92.4 & 0.290 & 0.291 & 0.284 \\
   \hline
\end{tabular}
\caption{\textit{Simulated coverage probability (in \%) and width of various 95\% confidence bands for the cumulative incidence function given a 45 years old female individual without pneumonia at time of hospital admission.}}
\label{tab:sim_results_C3}
\end{table}

In the following we describe the observations we made regarding the results of the simulation study. We focus on the comparison between the WB-MI confidence band ($\text{CB}^{*,\text{mi}}$) with its benchmark, the wild bootstrap confidence band for censoring-complete data ($\text{CB}^{*,\text{cc}}$), and the comparison between the (wild) bootstrap confidence band based on either multiple imputation ($\text{CB}^{*,\text{mi}}$) or based on IPCW ($\text{CB}^{\text{boot,ipcw}}$).

\begin{enumerate}
\item For the simulated settings the actual coverage probability of the corresponding methods turned out to lie in the following range:
\begin{itemize}
    \item $\text{CB}^{*,\text{cc}}$: $93.4\% - 95.4\%$;
    \item $\text{CB}^{*,\text{mi}}$: $93.4\% - 95.8\%$;
    \item $\text{CB}^{\text{boot,ipcw}}$ $91.8\% - 95.2\%$;
\end{itemize}
where the nominal coverage probability is $95\%$.
In particular,
compared to $\text{CB}^{\text{boot,ipcw}}$, the coverage probability of $\text{CB}^{*,\text{mi}}$ was always closer to the nominal level 95\%, except for one scenario.
Additionally, the actual coverage probability of $\text{CB}^{*,\text{mi}}$ was similar to the actual coverage probability of $\text{CB}^{*,\text{cc}}$, where in three out of the eight simulated settings the actual coverage probability of $\text{CB}^{*,\text{mi}}$ was closer to the nominal value and, in the other five cases, $\text{CB}^{*,\text{cc}}$ was more accurate.
\item The width of
    \begin{itemize}
    \item $\text{CB}^{*,\text{mi}}$ was comparable to the width of $\text{CB}^{*,\text{cc}}$;
    \item $\text{CB}^{*,\text{mi}}$ was greater than the width of $\text{CB}^{\text{boot,ipcw}}$.
    \end{itemize}
The latter relates to the relatively low coverage probabilities of $\text{CB}^{\text{boot,ipcw}}$.
\item As expected, an increase in sample size from 100 to 200 yielded a decrease in width of the confidence bands for all simulated settings and considered methods.
\item For the coverage probability of $\text{CB}^{*,\text{mi}}$ and $\text{CB}^{*,\text{cc}}$ under a setting with low/large numbers of imputations required  for $\text{CB}^{*,\text{mi}}$, i.e., the most/least favorable settings we found:
\begin{itemize}
    \item The most favorable setting described in \cref{sec:simulationStudy_Chp3} (strong censoring and baseline hazard rates $(\alpha^1_{01;0}, \alpha^1_{02;0})$) resulted in coverage probabilities of $94\%$ and $94.2\%$ for $\text{CB}^{*,\text{mi}}$ compared to the benchmark band $\text{CB}^{*,\text{cc}}$ with coverage probabilities $94.4\%$ and $93.4\%$ for $n=100$ and $n=200$, respectively;
    \item The least favorable setting described in \cref{sec:simulationStudy_Chp3} (light censoring and baseline hazard rates $(\alpha^2_{01;0}, \alpha^2_{02;0})$) led to coverage probabilities of $93.6\%$ for $\text{CB}^{*,\text{mi}}$ compared to the benchmark band $\text{CB}^{*,\text{cc}}$ with coverage probabilities $94.6\%$ and $93.8\%$ for $n=100$ and $n=200$, respectively.
\end{itemize}
Thus, the discrepancy between the coverage probability of $\text{CB}^{*,\text{mi}}$ compared to the coverage probability of its benchmark $\text{CB}^{*,\text{cc}}$
in the least favorable setting for a sample size of $100$ is with $1\%$ the largest of the four compared settings.
In contrast, the coverage probabilities of $\text{CB}^{*,\text{mi}}$ and $\text{CB}^{*,\text{cc}}$ are almost the same in the least favorable setting for a sample size of $200$ with a difference of only $0.2\%$. Hence, the effect of multiple imputation on the coverage of $\text{CB}^{*,\text{mi}}$ appears to be sensitive to the sample size in the most challenging setting. For the favorable setting, in which very few imputations are required, the absolute discrepancy between the coverage probabilities of $\text{CB}^{*,\text{mi}}$ and $\text{CB}^{*,\text{cc}}$ grows from $0.4\%$ to $0.8\%$ for $n=100$ growing to $200$.
\end{enumerate}
%Overall, we had expected a smaller difference in the simulation results between $\text{CB}^{*,\text{mi}}$ and $\text{CB}^{*,\text{cc}}$ for the most favorable compared to the least favorable setting.
%However, we do not observe this for the sample size 200.

In conclusion, the performance in terms of coverage probability of the WB-MI confidence band $\text{CB}^{*,\text{mi}}$ is better than the performance of the B-IPCW confidence band $\text{CB}^{\text{boot,ipcw}}$. Hence, out of the two considered approaches the WB-MI confidence band $\text{CB}^{*,\text{mi}}$ is the preferred choice in case of incomplete data.

%
%%
%%%
%%%%
%%%%%
%%%%%%
%%%%%%%
\section{Real Data Example}\label{sec:data_ex_C3}
%%%%%%%
%%%%%%
%%%%%
%%%%
%%%
%%
%

In this section, we illustrate the confidence bands obtained via the two approaches for handling incomplete data, multiple imputation and IPCW by applying them to a real data set. Both of the approaches incorporate generated information for the missing censoring times into the estimation of the cumulative incidence function. In contrast, the censoring-complete estimator is based on one realisation of the censoring times, such that it serves as a benchmark which is supposed to be retrieved by the other two estimators. Similarly, $CB_{1,n}^{*,\text{cc}}$ acts as the benchmark for both, ${CB}_{1,n}^{*,\text{mi}}$ and ${CB}_{1,n}^{\text{boot, ipcw}}$. Thus, we will contrast these confidence bands with each other and assess
%the WB-MI confidence band
${CB}_{1,n}^{*,\text{mi}}$ and
%the IPCW and bootstrap confidence band
$CB_{1,n}^{\text{boot, ipcw}}$ in terms of their deviation from the benchmark confidence band $CB_{1,n}^{*,\text{cc}}$.
%in terms of position, width and shape.

%
%%
%%%
%%%%
%%%%%
%%%%%%
%%%%%%%
\subsection{WB-MI Confidence Bands}\label{sec:data_ex_mi}
%%%%%%%
%%%%%%
%%%%%
%%%%
%%%
%%
%
%In this section, we illustrate the WB-MI confidence band introduced in \cref{subsec:imputation_estimates}. For this, w
We applied the proposed method to a data set obtained by merging the \textit{sir.adm} data set from the R-package \texttt{mvna} with the \textit{icu.pneu} data set from the R-package \texttt{kmi} by matching the patient ID.
That merged data set is censoring-complete.
These data were collected in the SIR 3 cohort study at Charit\'e university hospital Berlin, Germany.
It was concerned with the incidence of hospital-acquired infections in intensive care units (ICU).
Next to the right-censored event times, we included the patients' age, gender, and nosocomial pneumonia status (at baseline) in the Fine-Gray model.
We refer to \cite{SIR_results}, \cite{SIR_how_many}, \cite{SIR_beyersmann}, and \cite{SIR_wolkewitz} for more details and analyses of the data set.

We performed multiple imputation according to the three methods introduced in \cref{subsec:3imputations}. In particular, we used the Kaplan-Meier methodology, a Cox model and a uniform distribution to estimate the survival distribution of the censoring time in step \ref{item:imputation_step1} of \cref{WB-MI_scheme}. Here, we chose the uniform distribution as a suitable parametric distribution, because the Kaplan-Meier estimator of the censoring survival function exhibits a nearly constant slope for the given data set. For the particular choice of the uniform distribution, we fitted a straight line to the Kaplan-Meier estimate at time zero and at time 41.
%; this is the time by which 95\% of the individuals either have been censored or have experienced the event of interest.
Extrapolating this straight line resulted in the uniform distribution on the interval $[0,592]$. However, the duration of the study was 547 days only. Thus, we made a slight adjustment and took the uniform distribution on the interval $[0,547]$. We recall that the WB-MI confidence band was denoted by the superscript ``mi''. In view of the different estimation methods for the censoring survival distribution considered  in this section, we replace that superscript by either ``km'', ``cox'' or ``uni'' for the Kaplan-Meier methodology, the Cox model or the uniform distribution, respectively. The notation for the corresponding WB-MI confidence bands thus is $CB_{1,n}^{*,\text{km}}$, $CB_{1,n}^{*,\text{cox}}$, and $CB_{1,n}^{*,\text{uni}}$.

In the following, we will compare the three WB-MI confidence bands with each other. Thereby, we investigate the sensitivity of the proposed confidence band to the underlying multiple imputation technique for the data set at hand. Furthermore, we contrast either of $CB_{1,n}^{*,\text{mi}}$, $\text{mi}\in\{\text{km},\text{cox},\text{uni}\}$, with the benchmark confidence band $CB_{1,n}^{*,\text{cc}}$. All bands were computed for a 71 year old male individual with pneumonia on hospital admission; death is the event of interest. In case of the multiple imputation-based confidence bands $CB_{1,n}^{*,\text{mi}}$, $\text{mi}\in\{\text{km},\text{cox},\text{uni}\}$, the actual administrative censoring times for the individuals who experienced the competing event ``alive discharge from hospital'' provided by the data set were ignored and replaced by the imputed censoring times. For the multiple imputation according to the Kaplan-Meier methodology and the Cox model, the censoring times were generated using the \texttt{kmi} function provided by the \texttt{kmi} R-package \citep{kmi}, see also \cite{kmi_article}. Furthermore, the potential censoring times under the uniform assumption were generated manually by drawing from the conditional uniform distribution on the interval $[0,547]$, given that the censoring time exceeds the event time of the competing risk.

\begin{figure}%\label{fig:WB_variation}
    \centering
    \begin{subfigure}[H]{0.45\textwidth}
        \includegraphics[width=\textwidth]{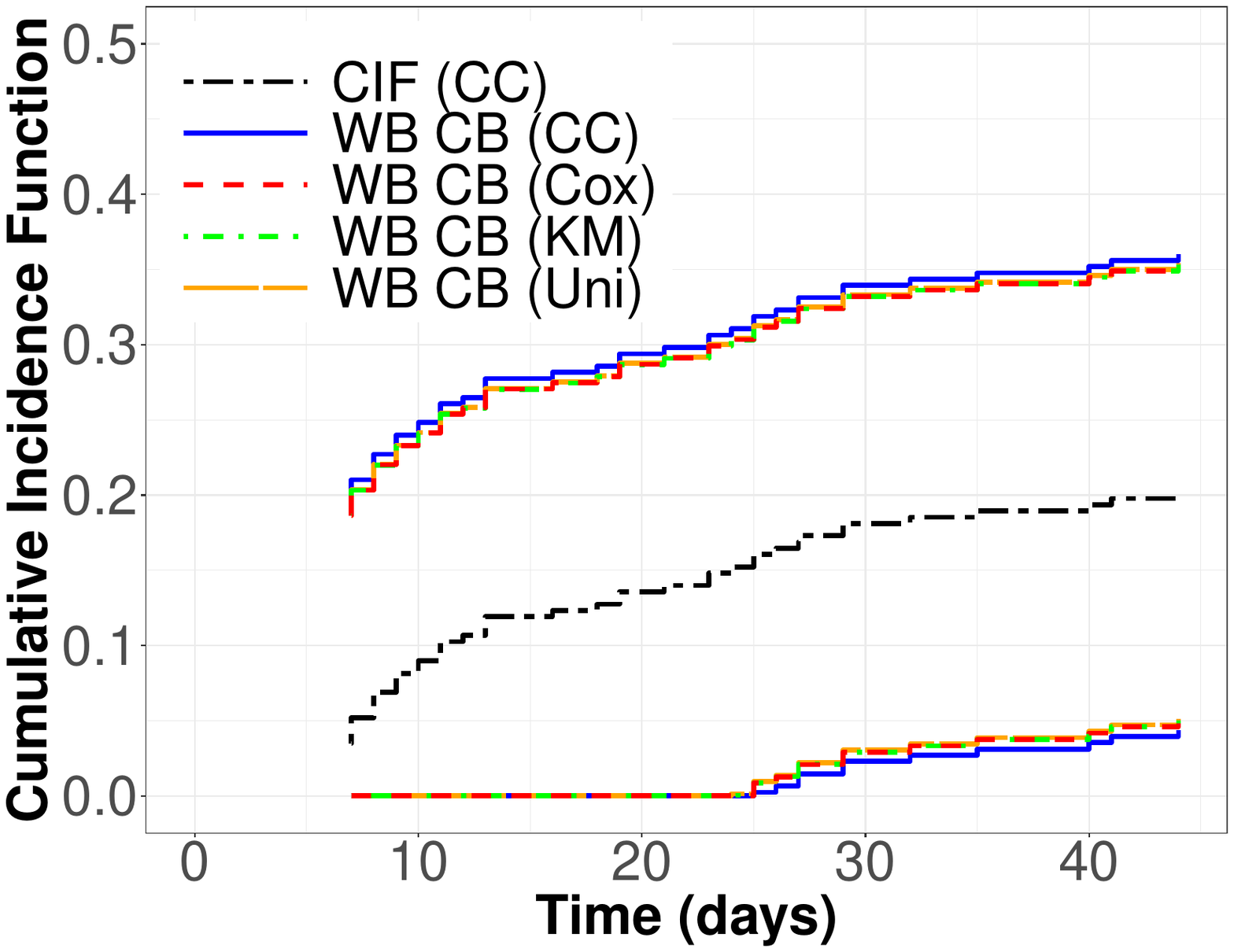}
        %\caption{The (overall) band estimates.}
        %\label{fig:CB_admCens_mean_imputation}
    \end{subfigure}
    \caption{\textit{The WB-MI confidence bands $CB_{1,n}^{*,\text{km}}$, $CB_{1,n}^{*,\text{cox}}$, and $CB_{1,n}^{*,\text{uni}}$, as well as the wild bootstrap confidence band $CB_{1,n}^{*,\text{cc}}$ including the estimated cumulative incidence function $\hat{F}_{1,n}^{\text{cc}}$ are depicted for a 71 year old male individual with pneumonia on hospital admission and death as the event of interest.}}\label{fig:WB_CB_adm_vs_imputed}
\end{figure}

In Figure~\ref{fig:WB_CB_adm_vs_imputed} the ${CB}_{1,n}^{*,\text{mi}}$ for all three imputation techniques $\text{mi} \in \{\text{km}, \text{cox},\text{uni}\}$, and the corresponding benchmark confidence band $CB_{1,n}^{*,\text{cc}}$ are plotted. The difference between all multiple imputation-based confidence bands is negligible. In particular, the boundaries after 44 days of $CB_{1,n}^{*,\text{km}}$, $CB_{1,n}^{*,\text{cox}}$, and $CB_{1,n}^{*,\text{uni}}$ are $(0.0502,0.3531), (0.0501,0.3532)$, and $(0.0514,0.3543)$, respectively. Thus, the eventual width of these bands is $0.3029, 0.3031$, and $0.3029$, respectively. We conclude that $CB_{1,n}^{*,\text{mi}}$ is not sensitive to the considered imputation techniques for the data set at hand. Moreover, the difference between the benchmark confidence band $CB_{1,n}^{*,\text{cc}}$ and any of the multiple imputation-based confidence band ${CB}_{1,n}^{*,\text{mi}}$, $\text{mi} \in \{\text{km}, \text{cox},\text{uni}\}$, is apparent but minor. In more detail, we observe that $CB_{1,n}^{*,\text{cc}}$ completely includes the multiple imputation-based confidence bands $CB_{1,n}^{*,\text{mi}}$ for any of the considered imputation techniques $\text{mi} \in \{\text{km}, \text{cox},\text{uni}\}$ with boundaries after 44 days equal to $(0.0437,0.3602)$ and a width of $0.3165$. Moreover, $CB_{1,n}^{*,\text{cox}}$ is the widest out of the three WB-MI confidence bands.
Finally, we choose the best multiple imputation-based confidence band as the one that is the closest to the benchmark confidence band $CB^{*,\text{cc}}_{1,n}$. As a consequence, $CB_{1,n}^{*,\text{cox}}$ is our preferred choice.
\subsection{WB-MI Confidence Bands vs.~B-IPCW Confidence Bands}\label{subsec:illustration_incomplete}
\begin{figure}%\label{fig:WB_variation}
    \centering
    \begin{subfigure}[H]{0.48\textwidth}
        \includegraphics[width=\textwidth]{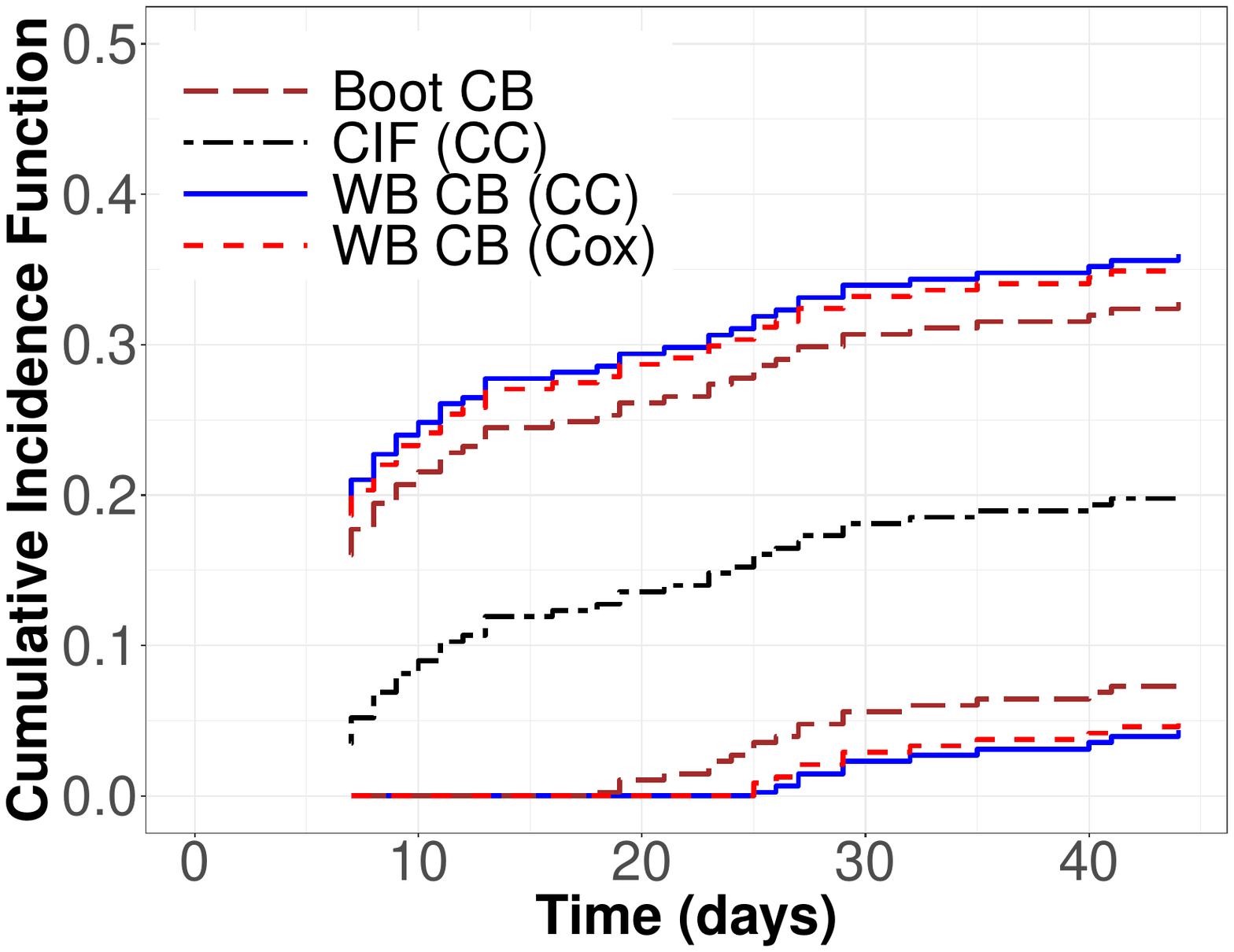}
        %\caption{$CB^{*,\text{PCC}}_{1,n}$ and $CB_{1,n}^{\text{boot}}$.}
        %\label{fig:CB_boot_Cox}
    \end{subfigure}
    \caption{\textit{Illustration and comparison of the B-IPCW confidence band $CB_{1,n}^{\text{boot, ipcw}}$ to the (multiple imputation and) wild bootstrap-based confidence bands $CB_{1,n}^{*,\text{cox}}$ and $CB^{*,\text{cc}}_{1,n}$.}}
    \label{fig:CB_boot_Cox}
\end{figure}

%%%%%%%
%%%%%%
%%%%%
%%%%
%%%
%%
%
In this section, we present the B-IPCW confidence band $CB_{1,n}^{\text{boot, ipcw}}$ and compare it to the best WB-MI confidence band  $CB_{1,n}^{*,\text{cox}}$ of \cref{sec:data_ex_mi}. The corresponding graphs can be found in Figure~\ref{fig:CB_boot_Cox}. Additionally, $CB_{1,n}^{*,\text{cc}}$ is depicted as the benchmark confidence band for both, $CB_{1,n}^{\text{boot, ipcw}}$ and $CB_{1,n}^{*,\text{cox}}$. We observe that $CB_{1,n}^{\text{boot, ipcw}}$ is clearly narrower than $CB_{1,n}^{*,\text{cox}}$. In other words, $CB_{1,n}^{\text{boot, ipcw}}$ is entirely contained in $CB_{1,n}^{*,\text{cox}}$ which in turn is completely overlapped by $CB_{1,n}^{*,\text{cc}}$.
Similarly as before, we assess the two competing confidence bands based on incomplete data, $CB_{1,n}^{*,\text{cox}}$ and $CB_{1,n}^{\text{boot, ipcw}}$, according to their deviation from the benchmark confidence band $CB^{*,\text{cc}}_{1,n}$. Hence, the WB-MI confidence band  $CB_{1,n}^{*,\text{cox}}$ is preferable over the B-IPCW confidence band $CB_{1,n}^{\text{boot, ipcw}}$.
%Thus, it is conceivable that the coverage probability of $CB_{1,n}^{\text{boot, ipcw}}$ is lower that the coverage probability of $CB_{1,n}^{*,\text{cox}}$ which is, in turn, lower than the coverage probability of $CB_{1,n}^{*,\text{cc}}$. Again, in alignment with the results of the simulation study, we assume that the coverage probability of $CB_{1,n}^{*,\text{cc}}$ is below its nominal value of $95\%$. As a consequence, the coverage probability of $CB_{1,n}^{*,\text{cox}}$ is closer to it nominal value of $95\%$ compared to the coverage probability of $CB_{1,n}^{\text{boot, ipcw}}$. Finally, under that assumption, we conclude that for the data set at hand the multiple imputation-based wild bootstrap confidence band  $CB_{1,n}^{*,\text{cox}}$ performs better than the bootstrap confidence band $CB_{1,n}^{\text{boot, ipcw}}$.
Additionally, in almost all settings of the simulation study the empirical coverage probabilities of $CB_{1,n}^{*,\text{cc}}$, $CB_{1,n}^{*,\text{cox}}$, and $CB_{1,n}^{\text{boot, ipcw}}$ were below the nominal value of $95\%$. This is another reason supporting $CB_{1,n}^{*,\text{cox}}$ as our preferred choice, because it is the widest of all confidence bands for incomplete data.

\iffalse
A related comparison has been conducted on the level of the regression coefficient by the authors of \cite{Fine-Gray}. In the simulation study of that paper, the authors contrasted the bias and the variance of the estimators of the regression coefficient obtained for censoring-complete data with those of the corresponding estimators obtained for incomplete data via IPCW.
\fi

%
%%
%%%
%%%%
%%%%%
%%%%%%
%%%%%%%
\section{Conclusion}
\label{sec:conclusion_Chp3}
%%%%%%%
%%%%%%
%%%%%
%%%%
%%%
%%
%
In this paper, we considered various confidence bands for the cumulative incidence function under the Fine-Gray model. We distinguished between three methods depending on the available information on the censoring times and on the chosen technique based on which censoring times, if missing, were accounted for. In particular, if a data set is censoring-complete, we considered the wild bootstrap confidence band as introduced in \cite{dietrich23} (Part~II) and summarized in \cref{CC_data}. Given incomplete data, we examined two techniques for handling the missing censoring times: multiple imputation and IPCW. In case of the former technique, which we named WB-MI, augmented data sets that are censoring-complete are created via multiple imputation. Based on each augmented data set, estimation of the cumulative incidence function and wild bootstrap resampling of that estimator can be performed as for censoring-complete data; see \cref{subsec:imputation_estimates}. The novelty of our WB-MI approach lies in the way the multiple imputation-based (wild bootstrap) estimators of the cumulative incidence function are combined in order to obtain confidence bands. %An alternative approach to deal with incomplete data is the so-called IPCW technique that can be used for the estimation of the cumulative incidence function when censoring times are missing. The resulting estimator can then be bootstrapped in order to obtain confidence bands; see \cref{subsec:incomplete_estimates}.

By means of simulations, we have studied the coverage probability as well as the  width of the considered confidence bands for the cumulative incidence function. The coverage probability of the WB-MI confidence band $CB_{1,n}^{*,\text{mi}}$ was found to be closer to the nominal value of $95\%$ than the coverage probability of the B-IPCW confidence band $CB_{1,n}^{\text{boot, ipcw}}$. Hence, our simulations suggest that the novel WB-MI confidence band $CB_{1,n}^{*,\text{mi}}$ is preferable over the B-IPCW confidence band $CB_{1,n}^{\text{boot, ipcw}}$. As expected, the benchmark confidence band $CB^{*,\text{cc}}_{1,n}$ based on censoring-complete data performs the best in terms of coverage probability. Moreover, the width of $CB_{1,n}^{*,\text{mi}}$ is comparable to the width of $CB^{*,\text{cc}}_{1,n}$.
We had expected that the uncertainty added due to the multiple imputation would increase the width of $CB_{1,n}^{*,\text{mi}}$ compared to the width of $CB^{*,\text{cc}}_{1,n}$. A possible explanation is that the between-imputation variance is minor; see the related comment on that matter in \cite{ruan_gray}.

We also illustrated the different confidence bands for the cumulative incidence function based on a real data example. In this context, we applied three multiple imputation techniques and compared the outcomes. We found that the WB-MI confidence band is not sensitive with respect to the chosen imputation method for this particular data set. Moreover, we observed that the B-IPCW confidence band $CB_{1,n}^{\text{boot, ipcw}}$ was considerable narrower than the WB-MI confidence band $CB_{1,n}^{*,\text{mi}}$, $\text{mi}\in\{\text{km},\text{cox},\text{uni}\}$ while the benchmark confidence band $CB^{*,\text{cc}}_{1,n}$ was the widest of all.
Because the wild bootstrap-based confidence band $CB_{1,n}^{*,\text{cox}}$ obtained via a Cox model in step \ref{item:imputation_step1} of \cref{WB-MI_scheme} exhibited the smallest distance to the benchmark confidence band $CB^{*,\text{cc}}_{1,n}$, this came out as our preferred band.

Proving the weak convergence and regularity conditions stated in \cref{assumpGaussian} and \cref{assumptions_C3} is part of future research. However, we would like to point out that all calculations in the arguments for the validity of \cref{assumptions_C3}, that were provided in \cref{subsec:3asymptotic},
%to verify the asymptotic correctness of the expectation and variance of the (wild bootstrap) overall estimator of the cumulative incidence function stated in \cref{thm:asymptotic_equivalence}.
%It is evident that all involved calculations
do not depend on the structure of the estimators, the specific choice of multiple imputation method, or the applied resampling method.
As a consequence, these calculations retain their validity beyond the particular choices we made and beyond the field of survival analysis.
In contrast, the crucial assumptions stated in \cref{assumptions_C3} naturally do depend on these specific choices.

%In future research, we intend to provide weak convergence results for the stochastic processes underlying the WB-MI confidence band to support the assumed asymptotic Gaussianity in \cref{subsec:3asymptotic}.
Another future goal is to construct a weighted and transformed version of the WB-MI confidence band. For this, a variance estimator for $\sqrt{n}\big(\phi (\overline{F}^{\text{mi}}_I) - \phi (F_1)\big)$, where $\phi$ denotes a transformation function, needs to be developed.

%
%%
%%%
%%%%
%%%%%
%%%%%%
%%%%%%%
%\section*{Appendix}
%\appendix
%%%%%%%
%%%%%%
%%%%%
%%%%
%%%
%%
%

%
%%
%%%
%%%%
%%%%%
%%%%%%
%%%%%%%
\section*{Appendix A: Proof and Remark}
%\section{Proof of \cref{thm:asymptotic_equivalence} }
%{Appendix: Arguments Underpinning the Proposed Imputation and Wild Bootstrap-based Confidence Band in Fine-Gray models}
\label{app:imputation_wbs}
%%%%%%%
%%%%%%\section{Appendix A: Proofs}\section*{Appendix A: Proofs and Remarks}
%%%%%
%%%%
%%%
%%
%

%\newcommand{\E}{\mathbb{E}}
%\newcommand{\Fhat}{\hat F_{1,n}^{\text{ic},m}}
%\newcommand{\Fstar}{\hat F_{1,n}^{*,\text{ic},m}}
%\newcommand{\FONEmisig}{\mathcal{F}_1^\text{mi}}
%\newcommand{\FbarM}{\overline{F}^{\text{mi}}_M}
%\newcommand{\FbarI}{\overline{F}^{\text{mi}}_I}

Below we will prove Theorem~\ref{thm:asymptotic_equivalence},  i.e., we will verify the statements \eqref{eq:statement_E} and \eqref{eq:statement_Var} about the asymptotic equality of the (conditional) expectations and variances of the assumed Gaussian limit processes of \eqref{process1} and \eqref{process2}. For simplicity, in this section we will refer to the cumulative incidence function $F_1(t|\textbf Z )$ and to any of its estimators without stating their argument $t|\textbf{Z}$, although they are considered pointwise in time and conditionally on $\textbf{Z}$.

As a preparation for the upcoming steps, from the definitions \eqref{Fbar-miM} and \eqref{Fbar*miI} we observe that
\begin{align}
\begin{split}
\label{preparation}
    \overline{F}^{*,\text{mi}}_{I} - \overline{F}^{\text{mi}}_{M} &= \frac{1}{I} \sum_{l=1}^{I} \hat{F}^{*,\text{ic},\pi(l)}_{1,n} - \frac{1}{M} \sum_{m=1}^{M}\hat{F}^{\text{ic},m}_{1,n}\\
    &=\frac{1}{I} \sum_{l=1}^{I} \sum_{m=1}^{M} X_{l,m}\hat{F}^{*,\text{ic},m}_{1,n} - \frac{1}{M } \sum_{m=1}^{M }\hat{F}^{\text{ic},m}_{1,n}\\
    &= \frac{1}{I} \sum_{l=1}^{I} \sum_{m=1}^{M} X_{l,m}(\hat{F}^{*,\text{ic},m}_{1,n} - \hat{F}^{\text{ic},m}_{1,n}) \\
    & \quad +\frac{1}{I} \sum_{l=1}^{I}  \sum_{m=1}^{M}(X_{l,m} - \frac{1}{M})\hat{F}^{\text{ic},m}_{1,n},
    \end{split}
\end{align}
where $X_{l,m} $ is the $m$-th element of the $M$-dimensional  random vector $\textbf{X}_{l} = (X_{l,1},\ldots , X_{l,M})$ that follows a multinomial distribution with parameters $N=1$ and $p = (p_1,\ldots,p_{M}) = (\frac{1}{M},\ldots, \frac{1}{M})$. In other words, $X_{l,m} = 1$, if during the $l$-th draw the augmented data set $\mathcal{D}^{m}$ has been selected ($\pi (l ) = m$), and $X_{l,m} = 0$ otherwise, $l\in\{1,\ldots , I\}$, so that
\begin{equation}
\label{exp-multinom}
\E(X_{l,m})=\frac{1}{M}, \quad m=1,\dots,M,
\end{equation}
for all $l=1,\ldots , I$.
We note that $X_{l,m} $ is independent of $\mathcal{F}_1^{\text{mi}}$ and of $\hat{F}_{1,n}^{ic,m}$, $m=1,\dots,M$, $l=1,\dots,I$.

\subsection*{Proof of \cref{thm:asymptotic_equivalence} }
%\textbf{Proof of Theorem~\ref{thm:asymptotic_equivalence}}.\\
In the following we assume that Assumptions~\ref{assumpGaussian} and \ref{assumptions_C3} hold.

%\subsection{Proof of Theorem~\ref{thm:asymptotic_equivalence}}

%\subsection{Asymptotic Equality of (Conditional) Expectations}

\smallskip
\noindent
\textbf{Proof of \eqref{eq:statement_E}}:\\
%Based on these assumptions, we first prove~\eqref{eq:statement_E}.
Beginning with the wild bootstrap-based expression, from \eqref{preparation} we see
\begin{align}\label{appendix_ch3_expectation}
\begin{split}
    \mathbb{E}\big( \sqrt{n}(\overline{F}^{*,\text{mi}}_{I} - \overline{F}^{\text{mi}}_{M}) |\mathcal{F}_1^{\text{mi}} \big)
    &=\mathbb{E}\big( \frac{1}{I} \sum_{l=1}^{I} \sum_{m=1}^{M} X_{l,m}\sqrt{n}(\hat{F}^{*,\text{ic},m}_{1,n} - \hat{F}^{\text{ic},m}_{1,n}) |\mathcal{F}_1^{\text{mi}} \big)\\
    &\quad + \mathbb{E}\big( \frac{1}{I} \sum_{l=1}^{I}  \sum_{m=1}^{M}\sqrt{n}(X_{l,m} - \frac{1}{M})\hat{F}^{\text{ic},m}_{1,n} |\mathcal{F}_1^{\text{mi}} \big)\\
    &= \frac{1}{I} \sum_{l=1}^{I} \sum_{m=1}^{M}\mathbb{E}\big( X_{l,m}\sqrt{n}(\hat{F}^{*,\text{ic},m}_{1,n} - \hat{F}^{\text{ic},m}_{1,n}) |\mathcal{F}_1^{\text{mi}} \big)\\
    &\quad + \frac{1}{I} \sum_{l=1}^{I}  \sum_{m=1}^{M}\sqrt{n}\hat{F}^{\text{ic},m}_{1,n}\mathbb{E}\big( X_{l,m} - \frac{1}{M}  \big)\\
    &= \frac{1}{I} \sum_{l=1}^{I} \sum_{m=1}^{M}\mathbb{E}\big( X_{l,m}\sqrt{n}(\hat{F}^{*,\text{ic},m}_{1,n} - \hat{F}^{\text{ic},m}_{1,n}) |\mathcal{F}_1^{\text{mi}} \big),
\end{split}
\end{align}
where in the second step we have applied that $\sigma(\hat{F}^{\text{ic},m}_{1,n})\subset\mathcal{F}_1^{\text{mi}}$ and that $X_{l,m}$ is independent of $\mathcal{F}_1^{\text{mi}}$. Additionally, in the third step we used \eqref{exp-multinom}.
%\eqref{appendix_ch3_expectation} that $\mathbb{E}\big( X_{l,m}   \big) = \frac{1}{M}$.
We continue with \eqref{appendix_ch3_expectation} and get
\begin{align}\label{appendix_ch3_expectation_2}
\begin{split}
    & \frac{1}{I} \sum_{l=1}^{I} \sum_{m=1}^{M}\mathbb{E}\big( X_{l,m}\big) \mathbb{E}\big(\sqrt{n}(\hat{F}^{*,\text{ic},m}_{1,n} - \hat{F}^{\text{ic},m}_{1,n}) |\mathcal{F}_1^{\text{mi}} \big)\\
    &= \frac{1}{M}\sum_{m=1}^{M} \mathbb{E}\big(\sqrt{n}(\hat{F}^{*,\text{ic},m}_{1,n} - \hat{F}^{\text{ic},m}_{1,n}) |\mathcal{F}_1^{\text{mi}} \big)\\
    &\stackrel{\mathbb{P}}{\longrightarrow} 0, \text { as } n\rightarrow \infty.
\end{split}
\end{align}
Note that we have used  on the left-hand side of the first equality of \eqref{appendix_ch3_expectation_2} that $X_{l,m}$ is independent of both $\sqrt{n}(\hat{F}^{*,\text{ic},m}_{1,n} - \hat{F}^{\text{ic},m}_{1,n})$ and $\mathcal{F}_1^{\text{mi}}$. Additionally, we made use of \eqref{ass:mom0*} in the second step of \eqref{appendix_ch3_expectation_2}. Combining \eqref{appendix_ch3_expectation} and \eqref{appendix_ch3_expectation_2}, we obtain
\begin{align}\label{eq:statement_E_1}
    \mathbb{E}\big( \sqrt{n}(\overline{F}^{*,\text{mi}}_{I} - \overline{F}^{\text{mi}}_{M}) |\mathcal{F}_1^{\text{mi}} \big) \stackrel{\mathbb{P}}{\longrightarrow} 0, \text { as } n\rightarrow \infty.
\end{align}
%\newpage
Furthermore, with \eqref{Fbar-miM}, \eqref{Fbar-miI} and the definition of the  $X_{l,m}$ we have
\begin{align}\label{eq:Form_FI__F1}
\begin{split}
    \overline{F}^{\text{mi}}_I - F_1 &= \frac{1}{I} \sum_{l=1}^I \hat{F}^{\text{ic},\pi (l)}_{1,n} -F_1\\
    &= \frac{1}{I} \sum_{l=1}^I \sum_{m=1}^M (X_{l,m} - \frac{1}{M}) \hat{F}^{\text{ic},m}_{1,n} + \overline{F}^{\text{mi}}_M - F_1.
\end{split}
\end{align}
Hence, we get
\begin{align}\label{appendix_ch3_expectation_3}
\begin{split}
    \mathbb{E} \big(\sqrt{n} (\overline{F}^{\text{mi}}_I - F_1) \big) &= \mathbb{E} \big( \frac{1}{I} \sum_{l=1}^I \sum_{m=1}^M \sqrt{n}(X_{l,m} - \frac{1}{M}) \hat{F}^{\text{ic},m}_{1,n} + \sqrt{n}(\overline{F}^{\text{mi}}_M - F_1) \big)\\
    &= \frac{1}{I} \sum_{l=1}^I \sum_{m=1}^M \mathbb{E} \big(  X_{l,m} - \frac{1}{M}\big)  \mathbb{E} \big(\sqrt{n}\hat{F}^{\text{ic},m}_{1,n}\big) + \mathbb{E} \big(\sqrt{n}(\overline{F}^{\text{mi}}_M - F_1 )\big)\\
    &= \mathbb{E} \big(\sqrt{n}(\overline{F}^{\text{mi}}_M - F_1) \big)\\
    &=\frac{1}{M}\sum_{m=1}^{M} \mathbb{E}\big (  \sqrt{n} (\hat{F}^{\text{ic},m}_{1,n} - F_1)\big ) \\
    &= \mathbb{E}\big (  \sqrt{n} (\hat{F}^{\text{ic},1}_{1,n} - F_1)\big )\longrightarrow 0, \text{ as } n\rightarrow \infty.
\end{split}
\end{align}
Here we used in the second step  that $X_{l,m}$ is independent of $\hat{F}^{\text{ic},m}_{1,n}$, and the third step holds because of \eqref{exp-multinom}.   In the fifth step, we have used that $\hat{F}^{\text{ic},1}_{1,n},\ldots ,\hat{F}^{\text{ic},M}_{1,n}$ are identically distributed, and the last step follows from \eqref{ass:mom0}.
Combining \eqref{eq:statement_E_1} with \eqref{appendix_ch3_expectation_3} yields \eqref{eq:statement_E}.
%\hfill\qedsymbol

%\subsection{Asymptotic Equality of (Conditional) Variances}
%\newpage
%\medskip
\noindent
\textbf{Proof of \eqref{eq:statement_Var}}:\\
For the upcoming considerations, we introduce the sigma-algebra $\mathcal{F}_2^{\text{mi}}$:
$$\mathcal{F}_2^{\text{mi}} = \sigma \{\mathcal{D}^m, G_i  \mathbbm{1}\{T_i \leq C_i, \epsilon_i = 1, T_i \leq \tau\}, m=1,\ldots , M , i=1,\ldots , n\}.$$
Note that $X_{l,m}$ is independent of $\mathcal{F}_2^{\text{mi}}$, $m=1,\dots,M, l=1,\ldots,I$.
%\newpage
Additionally, it will turn out to be fruitful to use the following formulation
\begin{align}
\label{diffFbars}
\begin{split}
    \overline{F}^{*,\text{mi}}_{I} - \overline{F}^{\text{mi}}_{M} &= \frac{1}{I} \sum_{l=1}^{I} \hat{F}^{*,\text{ic},\pi(l)}_{1,n} - \frac{1}{M} \sum_{m=1}^{M}\hat{F}^{\text{ic},m}_{1,n}\\
    &=\frac{1}{I} \sum_{l=1}^{I} \sum_{m=1}^{M} X_{l,m}\hat{F}^{*,\text{ic},m}_{1,n} - \frac{1}{M } \sum_{m=1}^{M }\hat{F}^{\text{ic},m}_{1,n}\\
    &= \frac{1}{I} \sum_{l=1}^{I} \sum_{m=1}^{M} (X_{l,m} - \frac{1}{M})\hat{F}^{*,\text{ic},m}_{1,n} \\
    & \quad + \frac{1}{M } \sum_{m=1}^{M } (\hat{F}^{*,\text{ic},m}_{1,n} - \hat{F}^{\text{ic},m}_{1,n}).
    \end{split}
\end{align}
With \eqref{diffFbars}, we get
\begin{align}\label{eq:appendix_Var_star_1}
\begin{split}
    &\text{Var}\big (\overline{F}^{*,\text{mi}}_{I} - \overline{F}^{\text{mi}}_{M}| \mathcal{F}_1^{\text{mi}} \big)\\
    &=\text{Var}\big ( \frac{1}{I} \sum_{l=1}^{I} \sum_{m=1}^{M} (X_{l,m} - \frac{1}{M})\hat{F}^{*,\text{ic},m}_{1,n} + \frac{1}{M } \sum_{m=1}^{M } (\hat{F}^{*,\text{ic},m}_{1,n} - \hat{F}^{\text{ic},m}_{1,n})  | \mathcal{F}_1^{\text{mi}} \big)\\
    &= \mathbb{E}\big( \text{Var}\big ( \frac{1}{I} \sum_{l=1}^{I} \sum_{m=1}^{M} (X_{l,m} - \frac{1}{M})\hat{F}^{*,\text{ic},m}_{1,n} + \frac{1}{M } \sum_{m=1}^{M } (\hat{F}^{*,\text{ic},m}_{1,n} - \hat{F}^{\text{ic},m}_{1,n})  | \mathcal{F}_2^{\text{mi}} \big) |  \mathcal{F}_1^{\text{mi}} \big)\\
    & \quad + \text{Var}\big( \mathbb{E}\big ( \frac{1}{I} \sum_{l=1}^{I} \sum_{m=1}^{M} (X_{l,m} - \frac{1}{M})\hat{F}^{*,\text{ic},m}_{1,n} + \frac{1}{M } \sum_{m=1}^{M } (\hat{F}^{*,\text{ic},m}_{1,n} - \hat{F}^{\text{ic},m}_{1,n})   | \mathcal{F}_2^{\text{mi}} \big) |  \mathcal{F}_1^{\text{mi}} \big)\\
    &= \mathbb{E}\big( \frac{1}{I^2} \sum_{l_1, l_2 =1}^{I} \sum_{m_1, m_2 =1}^{M} \text{Cov}\big (   X_{l_1,m_1}, X_{l_2,m_2} \big) \hat{F}^{*,\text{ic},m_1}_{1,n} \hat{F}^{*,\text{ic},m_1}_{1,n} |  \mathcal{F}_1^{\text{mi}} \big)\\
    & \quad + \text{Var}\big( \frac{1}{I} \sum_{l=1}^{I}\sum_{m=1}^{M} \mathbb{E}\big (   X_{l,m} - \frac{1}{M} \big) \hat{F}^{*,\text{ic},m}_{1,n} + \frac{1}{M } \sum_{m=1}^{M } (\hat{F}^{*,\text{ic},m}_{1,n} - \hat{F}^{\text{ic},m}_{1,n}) |  \mathcal{F}_1^{\text{mi}} \big)\\
    &= \mathbb{E}\big( \frac{1}{I^2} \sum_{l_1, l_2 =1}^{I} \sum_{m_1, m_2 =1}^{M} \text{Cov}\big (   X_{l_1,m_1}, X_{l_2,m_2} \big) \hat{F}^{*,\text{ic},m_1}_{1,n} \hat{F}^{*,\text{ic},m_1}_{1,n} |  \mathcal{F}_1^{\text{mi}} \big)\\
    & \quad + \text{Var}\big(  \frac{1}{M } \sum_{m=1}^{M } (\hat{F}^{*,\text{ic},m}_{1,n} - \hat{F}^{\text{ic},m}_{1,n}) |  \mathcal{F}_1^{\text{mi}} \big),
\end{split}
\end{align}
where in the third step we have used that $\sigma(\hat{F}^{\text{ic},m}_{1,n})\subset\mathcal{F}_2^{\text{mi}}$ and $\sigma(\hat{F}^{*,\text{ic},m}_{1,n})\subset\mathcal{F}_2^{\text{mi}}$ holds for all $m=1,\ldots , M$. Additionally, we exploited in that step that $X_{l,m}$ is independent of $\mathcal{F}_2^{\text{mi}}$. Moreover, in the fourth step we used \eqref{exp-multinom}. We continue with \eqref{eq:appendix_Var_star_1} and obtain
\begin{align}\label{eq:appendix_Var_star_2}
\begin{split}
&\mathbb{E}\big( \frac{1}{I^2} \sum_{l_1, l_2 =1}^{I} \sum_{m_1, m_2 =1}^{M} \text{Cov}\big (   X_{l_1,m_1}, X_{l_2,m_2} \big) \hat{F}^{*,\text{ic},m_1}_{1,n} \hat{F}^{*,\text{ic},m_1}_{1,n} |  \mathcal{F}_1^{\text{mi}} \big)\\
& \quad + \text{Var}\big(  \frac{1}{M } \sum_{m=1}^{M } (\hat{F}^{*,\text{ic},m}_{1,n} - \hat{F}^{\text{ic},m}_{1,n}) |  \mathcal{F}_1^{\text{mi}} \big)\\
    &=\mathbb{E}\big( \frac{1}{I} \big [
     \sum_{m =1}^{M} \text{Var}\big (   X_{1,1} \big)  (\hat{F}^{*,\text{ic},m}_{1,n})^2 \\
     & \quad \quad +  \sum_{m_1 \neq m_2 } \text{Cov}\big (   X_{1,m_1}, X_{1,m_2} \big)
    \hat{F}^{*,\text{ic},m_1}_{1,n} \hat{F}^{*,\text{ic},m_2}_{1,n}
    \big ] |  \mathcal{F}_1^{\text{mi}} \big)\\
    &\quad +\text{Var}\big( \frac{1}{M } \sum_{m=1}^{M } (\hat{F}^{*,\text{ic},m}_{1,n} - \hat{F}^{\text{ic},m}_{1,n}) |  \mathcal{F}_1^{\text{mi}} \big)\\
    %& \quad + \text{Var}\big( \frac{1}{I} \sum_{l=1}^{I}\sum_{m=1}^{M} \mathbb{E}\big (   X_{l,m} - \frac{1}{M} \big) \hat{F}^{*,\text{ic},m}_{1,n} |  \mathcal{F}^{\text{ic}}(\tau) \big)\\
    &=\mathbb{E}\big( \frac{1}{I} \big [
      \frac{M-1}{M^2} \sum_{m=1}^M (\hat{F}^{*,\text{ic},m}_{1,n})^2 -    \frac{1}{M^2} \sum_{m_1 \neq m_2} \hat{F}^{*,\text{ic},m_1}_{1,n} \hat{F}^{*,\text{ic},m_2}_{1,n}
    \big ] |  \mathcal{F}_1^{\text{mi}} \big)\\
    &\quad +\frac{1}{M^2 } \big [\sum_{m=1}^{M }\text{Var}\big(  \hat{F}^{*,\text{ic},m}_{1,n} |  \mathcal{F}_1^{\text{mi}} \big)
    + \sum_{m_1\neq m_2}\text{Cov}\big(  \hat{F}^{*,\text{ic},m_1}_{1,n}, \hat{F}^{*,\text{ic},m_2}_{1,n} |  \mathcal{F}_1^{\text{mi}} \big)
    \big ] ,
\end{split}
\end{align}
where for the first equality we have used that $\textbf{X}_{1}, \dots, \textbf{X}_{I}$ are independent and identically distributed, and that $\text{Var}\big( X_{1,m} \big)$ is identical for all $m=1,\ldots,M$. In particular, the independence of $\textbf{X}_{1}, \dots, \textbf{X}_{I}$ implies that $\text{Cov}\big ( X_{l_1,m_1}, X_{l_2,m_2} \big ) = 0 $ for $l_1 \neq l_2$. Furthermore, the second equality holds because $\text{Var}\big( X_{1,1} \big) = \frac{M-1}{M^2}$, $\text{Cov}\big( X_{1,m_1}, X_{1,m_2} \big) = -\frac{1}{M^2}$, and $\sigma(\hat{F}^{\text{ic},m}_{1,n})\subset\mathcal{F}_1^{\text{mi}}$. We proceed from the right-hand side of \eqref{eq:appendix_Var_star_2}.
%\newpage

Some elementary properties yield
\begin{align}\label{eq:appendix_Var_star_2.2}
\begin{split}
%\begin{multline}\label{eq:appendix_Var_star_2.2}
&\mathbb{E}\big( \frac{1}{I} \big [
      \frac{M-1}{M^2} \sum_{m=1}^M (\hat{F}^{*,\text{ic},m}_{1,n})^2 -    \frac{1}{M^2} \sum_{m_1 \neq m_2} \hat{F}^{*,\text{ic},m_1}_{1,n} \hat{F}^{*,\text{ic},m_2}_{1,n}
    \big ] |  \mathcal{F}_1^{\text{mi}} \big)\\
    &\quad +\frac{1}{M^2 } \big [\sum_{m=1}^{M }\text{Var}\big(  \hat{F}^{*,\text{ic},m}_{1,n} |  \mathcal{F}_1^{\text{mi}} \big)
    + \sum_{m_1\neq m_2}\text{Cov}\big(  \hat{F}^{*,\text{ic},m_1}_{1,n}, \hat{F}^{*,\text{ic},m_2}_{1,n} |  \mathcal{F}_1^{\text{mi}} \big)
    \big ] \\
     &=\frac{1}{I} \big [  \frac{M-1}{M^2} \sum_{m=1}^M \text{Var}\big(
       \hat{F}^{*,\text{ic},m}_{1,n}|  \mathcal{F}_1^{\text{mi}} \big)
       + \frac{M-1}{M^2} \sum_{m=1}^M \mathbb{E}\big(
       \hat{F}^{*,\text{ic},m}_{1,n}|  \mathcal{F}_1^{\text{mi}} \big)^2
       \\
       & \quad -  \frac{1}{M^2} \sum_{m_1 \neq m_2} \text{Cov}\big(  \hat{F}^{*,\text{ic},m_1}_{1,n}, \hat{F}^{*,\text{ic},m_2}_{1,n}
     |  \mathcal{F}_1^{\text{mi}} \big)
     -  \frac{1}{M^2} \sum_{m_1 \neq m_2} \mathbb{E}\big(  \hat{F}^{*,\text{ic},m_1}_{1,n} |  \mathcal{F}_1^{\text{mi}} \big) \mathbb{E}\big(\hat{F}^{*,\text{ic},m_2}_{1,n}
     |  \mathcal{F}_1^{\text{mi}} \big) \big ]\\
    &\quad +\frac{1}{M^2 } \big [\sum_{m=1}^{M } \text{Var}\big(  \hat{F}^{*,\text{ic},m}_{1,n} |  \mathcal{F}_1^{\text{mi}} \big)
   + \sum_{m_1\neq m_2} \text{Cov}\big(  \hat{F}^{*,\text{ic},m_1}_{1,n} ,\hat{F}^{*,\text{ic},m_2}_{1,n} |  \mathcal{F}_1^{\text{mi}} \big)
    \big ]     \\
    &= \frac{I + M -1}{ M^2 I} \sum_{m=1}^M \text{Var}\big(
    \hat{F}^{*,\text{ic},m}_{1,n}|  \mathcal{F}_1^{\text{mi}} \big)
    +  \frac{I-1}{ M^2 I} \sum_{m_1 \neq m_2} \text{Cov}\big(  \hat{F}^{*,\text{ic},m_1}_{1,n} ,\hat{F}^{*,\text{ic},m_2}_{1,n} |  \mathcal{F}_1^{\text{mi}} \big)\\
    & \quad  + \frac{M-1}{M^2 I} \sum_{m=1}^M \mathbb{E}\big(
       \hat{F}^{*,\text{ic},m}_{1,n}|  \mathcal{F}_1^{\text{mi}} \big)^2
     -  \frac{1}{M^2I} \sum_{m_1 \neq m_2} \mathbb{E}\big(  \hat{F}^{*,\text{ic},m_1}_{1,n} |  \mathcal{F}_1^{\text{mi}} \big) \mathbb{E}\big(\hat{F}^{*,\text{ic},m_2}_{1,n}
     |  \mathcal{F}_1^{\text{mi}} \big).
\end{split}
%\end{multline}
\end{align}
Thus, from \eqref{eq:appendix_Var_star_1}, \eqref{eq:appendix_Var_star_2}, \eqref{eq:appendix_Var_star_2.2} we have
\begin{align}\label{eq:appendix_Var_star_3}
\begin{split}
    \text{Var}\big (\overline{F}^{*,\text{mi}}_{I} - \overline{F}^{\text{mi}}_{M}| \mathcal{F}_1^{\text{mi}} \big) &= \frac{I + M -1}{ M^2 I} \sum_{m=1}^M \text{Var}\big(
    \hat{F}^{*,\text{ic},m}_{1,n}|  \mathcal{F}_1^{\text{mi}} \big)
    \\
    & \quad+  \frac{I-1}{ M^2 I} \sum_{m_1 \neq m_2} \text{Cov}\big(  \hat{F}^{*,\text{ic},m_1}_{1,n} ,\hat{F}^{*,\text{ic},m_2}_{1,n} |  \mathcal{F}_1^{\text{mi}} \big)\\
    & \quad  + \frac{1}{M I} \sum_{m=1}^M \mathbb{E}\big(
       \hat{F}^{*,\text{ic},m}_{1,n}|  \mathcal{F}_1^{\text{mi}} \big)^2
     -  \frac{1}{M^2I} \big[\sum_{m=1}^M \mathbb{E}\big(  \hat{F}^{*,\text{ic},m}_{1,n} |  \mathcal{F}_1^{\text{mi}} \big) \big]^2.
\end{split}
\end{align}
We will now consider the terms on the right-hand sides of \eqref{eq:appendix_Var_star_3}, term by term.

%{\ }\\
%{\bf {Consideration of the first term:}}\\
%We start with the expression for the wild bootstrap.
For the first term on the right-hand side of \eqref{eq:appendix_Var_star_3} it follows from \eqref{ass:mom21*} %and \eqref{ass:mom_eps3}
that
\begin{align}
\label{eq:comparison11}
\begin{split}
 & \frac{I+M-1}{M^2I} \sum_{m=1}^M \text{Var}(\Fstar | \FONEmisig) \\
 & = \frac{I+M-1}{M^2I} \sum_{m=1}^M \text{Var}(\Fhat) + \frac{I+M-1}{MI} o_p(n^{-1}) \\
 & = \frac{I+M-1}{MI} \cdot \text{Var}(\hat{F}^{\text{ic},1}_{1,n})+ \frac{I+M-1}{MI} \cdot o_p(n^{-1}) ,
 \end{split}
\end{align}
where in the second step we have used that $\hat{F}^{\text{ic},1}_{1,n},\ldots ,\hat{F}^{\text{ic},M}_{1,n}$ are identically distributed.

%Note that the term on the right-hand side of the second equality equals the first term of~\eqref{eq:appendix_Var_2} plus a remainder term.\\
%{\ }\\
%{\bf {Consideration of the second term:}}\\
%Let us again take a look at the wild bootstrap.
With \eqref{ass:mom21*}, we find for the second term on the right-hand side of \eqref{eq:appendix_Var_star_3}
\begin{align}
\label{eq:comparison21}
\begin{split}
 & \frac{I-1}{M^2I} \sum_{m_1 \neq m_2} \text{Cov}( \hat F_{1,n}^{*,\text{ic},m_1}, \hat F_{1,n}^{*,\text{ic},m_2} | \FONEmisig) \\
 & = \frac{(M-1)(I-1)}{MI} \cdot \text{Cov}(\hat F_{1,n}^{\text{ic},1},\hat F_{1,n}^{\text{ic},2}) + \frac{(M-1)(I-1)}{MI} \cdot o_p(n^{-1}),
\end{split}
\end{align}
where we have used that $\hat{F}^{\text{ic},1}_{1,n},\ldots ,\hat{F}^{\text{ic},M}_{1,n}$ are identically distributed.

%Again, the term on the right-hand side equals the second term of~\eqref{eq:appendix_Var_2} plus a remainder term.\\
%{\ }\\
%{\bf {Consideration of the remaining terms:}}\\
%We will show that the last two terms of \eqref{eq:appendix_Var_star_3} are negligible with a sufficiently high rate.
Let us now consider the last two terms of the right-hand side of \eqref{eq:appendix_Var_star_3}. With Assumption~\eqref{ass:mom0*} we rewrite these two terms as follows
\begin{align*}
%\label{eq:comparison31}
%\begin{split}
& \frac1{MI}\sum_{m=1}^M \E(\Fstar| \FONEmisig)^2 - \frac1{M^2I}\big[\sum_{m=1}^M \E(\Fstar| \FONEmisig) \big]^2
\\
& = \frac1{MI}\sum_{m=1}^M \big[ \Fhat + o_p(n^{-1}) \big]^2  - \frac1{M^2I}\big[\sum_{m=1}^M \{\Fhat + o_p(n^{-1}) \} \big]^2
\\
& = \frac1{MI}\sum_{m=1}^M \big[ (\Fhat)^2 + 2 \Fhat o_p(n^{-1}) + o_p(n^{-2}) \big] \\
& \quad - \frac1{M^2I}\big[\sum_{m=1}^M \{\Fhat - F_1 \} + M \{ F_1 + o_p(n^{-1}) \} \big]^2
\\
&
= \frac1{MI}\sum_{m=1}^M \big[ (\Fhat)^2 - F_1^2 \big] +  \frac2{MI}\sum_{m=1}^M \big[ \Fhat - F_1 \big]  o_p(n^{-1}) \\
&\quad + \frac1I F_1^2 + \frac2I F_1  o_p(n^{-1}) + o_p(n^{-2})  \\
& \quad - \frac1{M^2I}\big[\sum_{m=1}^M \{\Fhat - F_1 \} \big]^2 - \frac2{M I} \sum_{m=1}^M \{\Fhat - F_1 \}   \{ F_1 + o_p(n^{-1}) \}  - \frac1I  \{ F_1 + o_p(n^{-1}) \}^2
\\
&
= \frac1{MI}\sum_{m=1}^M \big[ (\Fhat)^2 - F_1^2 \big] -  \frac{2 F_1 + o_p(n^{-1})}{MI}\sum_{m=1}^M \big[ \Fhat - F_1 \big]
- \frac1{M^2I}\big[\sum_{m=1}^M \{\Fhat - F_1 \} \big]^2 \\
& \quad + o_p(n^{-2} I^{-1}) + o_p(n^{-2}).
 %\end{split}
\end{align*}
Applications of Assumptions~\eqref{ass:mom_eps2} and~\eqref{ass:mom_eps3}
reveal the following rates of the terms in the previous display:
\begin{align}\label{eq:O_p_3rd_terms}
\begin{split}
    & O_p(M^{-1/2} I^{-1}) + O_p(n^{-1/2} I^{-1}) + O_p(M^{-1/2} I^{-1}) + O_p(n^{-1/2} I^{-1}) + O_p(M^{-1}I^{-1}) \\
    & + O_p(M^{-1/2}n^{-1/2} I^{-1})
    + O_p(n^{-1} I^{-1}) + o_p(n^{-2} I^{-1}) + o_p(n^{-2}) \\
    & = O_p(M^{-1/2} I^{-1}) + O_p(n^{-1/2} I^{-1})
     + o_p(n^{-2}) .
\end{split}
\end{align}
Hence, we obtain %for the remaining terms of \eqref{eq:appendix_Var_star_3} that
\begin{align}\label{eq:comparison31}
\begin{split}
    &\frac1{MI}\sum_{m=1}^M \E(\Fstar| \FONEmisig)^2 - \frac1{M^2I}\big[\sum_{m=1}^M \E(\Fstar| \FONEmisig) \big]^2 \\
    &= O_p(M^{-1/2} I^{-1}) + O_p(n^{-1/2} I^{-1})
     + o_p(n^{-2}) .
\end{split}
\end{align}
\quad \\
Combining \eqref{eq:appendix_Var_star_3}, \eqref{eq:comparison11}, \eqref{eq:comparison21}, and \eqref{eq:comparison31} as well as applying Assumptions~\eqref{ass:mom1}, \eqref{ass:mom21}, \eqref{ass:I}, and~\eqref{ass:M}, yields
\begin{align}\label{eq:Var_F_bar_star}
\begin{split}
    \text{Var}\big( \sqrt{n}(
    \hat{F}^{*,\text{ic},m}_{1,n}|  \mathcal{F}_1^{\text{mi}}) \big)
    &=  n\big [ \frac{I+M-1}{MI} \{ \text{Var}(\hat{F}^{\text{ic},1}_{1,n}) +  o_p(n^{-1})\} \\
    &\quad + \frac{(M-1)(I-1)}{MI} \{ \text{Cov}(\hat F_{1,n}^{\text{ic},1},\hat F_{1,n}^{\text{ic},2}) +  o_p(n^{-1}) \}\\
    &\quad  + O_p(M^{-1/2} I^{-1}) + O_p(n^{-1/2} I^{-1})
     + o_p(n^{-2})  \big]\\
     &= \frac{I+M-1}{MI} \big[\text{V} + o(1) +  o_p(1)\big] \\
    &\quad + \frac{(M-1)(I-1)}{MI} \big[ \text{C} + o(1) +  o_p(1) \big ]\\
    &\quad  + o_p(1)  + o_p(n^{-1}) \\
    & \stackrel{\mathbb P}{\longrightarrow} C, \text{ as } n,M(n),I(n) \rightarrow \infty.
\end{split}
\end{align}
Next, we consider $\text{Var}\big (\overline{F}^{\text{mi}}_{I} - F_1 \big)$. For this, we resort to \eqref{eq:Form_FI__F1} and obtain
\begin{align}\label{eq:appendix_Var_1}
    \begin{split}
        \text{Var}\big (\overline{F}^{\text{mi}}_{I} - F_1 \big)
        &= \text{Var}\big ( \frac{1}{I} \sum_{l=1}^I \sum_{m=1}^M (X_{l,m} - \frac{1}{M}) \hat{F}^{\text{ic},m}_{1,n} + \overline{F}^{\text{mi}}_M - F_1 \big)\\
        &= \mathbb{E}\big( \text{Var}\big ( \frac{1}{I} \sum_{l=1}^I \sum_{m=1}^M (X_{l,m} - \frac{1}{M}) \hat{F}^{\text{ic},m}_{1,n} + \overline{F}^{\text{mi}}_M - F_1 | \mathcal{F}_1^{\text{mi}} \big)  \big)\\
    & \quad + \text{Var}\big( \mathbb{E}\big (  \frac{1}{I} \sum_{l=1}^I \sum_{m=1}^M (X_{l,m} - \frac{1}{M}) \hat{F}^{\text{ic},m}_{1,n} + \overline{F}^{\text{mi}}_M - F_1  | \mathcal{F}_1^{\text{mi}} \big)\big)\\
    &= \mathbb{E}\big( \frac{1}{I^2} \sum_{l_1, l_2 =1}^{I} \sum_{m_1, m_2 =1}^{M} \text{Cov}\big (   X_{l_1,m_1}, X_{l_2,m_2} \big) \hat{F}^{\text{ic},m_1}_{1,n} \hat{F}^{\text{ic},m_2}_{1,n} \big)\\
    & \quad + \text{Var}\big( \overline{F}^{\text{mi}}_M - F_1 \big ) ,
    \end{split}
\end{align}
where in the third step we have used that $\sigma(\hat{F}^{\text{ic},m}_{1,n})\subset\mathcal{F}_1^{\text{mi}}$, $\sigma(\overline{F}^{\text{mi}}_{M})\subset\mathcal{F}_1^{\text{mi}}$, and $\sigma({F}_1)\subset\mathcal{F}_1^{\text{mi}}$ holds for all $m=1,\ldots , M$. Additionally, we applied in that step that $X_{l,m}$ is independent of $\mathcal{F}_1^{\text{mi}}$ and \eqref{exp-multinom}.

Note that all arguments used to justify \eqref{eq:appendix_Var_star_2} and to gain \eqref{eq:appendix_Var_star_2.2} hold analogously for the right-hand side of the third equation of \eqref{eq:appendix_Var_1}. Thus, we obtain the following counterpart of \eqref{eq:appendix_Var_star_3}:
\begin{align}\label{eq:appendix_Var_2}
\begin{split}
    \text{Var}\big (\overline{F}^{\text{mi}}_{I} - F_1\big) &= \frac{I + M -1}{ M^2 I} \sum_{m=1}^M \text{Var}\big(
       \hat{F}^{\text{ic},m}_{1,n} \big)\\
    &\quad +  \frac{I-1}{ M^2 I} \sum_{m_1 \neq m_2} \text{Cov}\big(  \hat{F}^{\text{ic},m_1}_{1,n}, \hat{F}^{\text{ic},m_2}_{1,n}  \big)\\
    & \quad + \frac{1}{MI} \sum_{m=1}^M \mathbb{E}(\hat{F}^{\text{ic},m}_{1,n})^2 - \frac{1}{M^2I} \big [\sum_{m=1}^M \mathbb{E}\big(  \hat{F}^{\text{ic},m}_{1,n} \big) \big]^2
    \\
    & = \frac{I + M -1}{ M I} \text{Var}\big(
       \hat{F}^{\text{ic},1}_{1,n}
       \big) +  \frac{(M-1)(I-1)}{ M I}  \text{Cov}\big(  \hat{F}^{\text{ic},1}_{1,n}, \hat{F}^{\text{ic},2}_{1,n}  \big) \\
       & \quad + \frac1I \mathbb{E}(\hat{F}^{\text{ic},1}_{1,n})^2
       - \frac1I \mathbb{E}(\hat{F}^{\text{ic},1}_{1,n})^2.
\end{split}
\end{align}
In the second step, we have used that $\hat{F}^{\text{ic},1}_{1,n},\ldots ,\hat{F}^{\text{ic},M}_{1,n}$ are identically distributed. As can be taken from the right-hand side of the second equality, the terms related to the first moments vanish.
%\newpage

Eventually, it follows from \eqref{eq:appendix_Var_2} with \eqref{ass:I} and \eqref{ass:M} that
\begin{align}\label{eq:Var_F_bar}
\begin{split}
    \text{Var}\big (\sqrt{n}(\overline{F}^{\text{mi}}_{I} - F_1)\big) &= n\big[\frac{I + M -1}{ M I} \text{Var}\big(
       \hat{F}^{\text{ic},1}_{1,n}
       \big) +  \frac{(M-1)(I-1)}{ M I}  \text{Cov}\big(  \hat{F}^{\text{ic},1}_{1,n}, \hat{F}^{\text{ic},2}_{1,n}  \big) \big]\\
      &= \frac{I + M -1}{ M I} \big[ V + o(1) \big ]
      + \frac{(M-1)(I-1)}{ M I}  \big[ C + o(1)  \big ]\\
      &\longrightarrow C, \text{ as } n,M(n),I(n) \rightarrow \infty.
\end{split}
\end{align}
Finally, combining \eqref{eq:Var_F_bar_star} and \eqref{eq:Var_F_bar}, yields \eqref{eq:statement_Var}.
%The assumed convergence rates in~\eqref{ass:I} and~\eqref{ass:M} imply that this is of order $o_p(n^{-1})$.
\hfill\qedsymbol
\\
{\ }\\
{\bf {Final remark on~\cref{thm:asymptotic_equivalence} and its proof:}}\\
The term $O_p(n^{-1/2} I^{-1})$ in \eqref{eq:O_p_3rd_terms} revealed that there is a trade-off between the assumed biases of $\Fhat$ and $\Fstar$ and the convergence rate of $I=I(n)$ as $n\to\infty$.
If smaller biases would lead to a term like $O_p(n^{-1/2 - \kappa} I^{-1})$ with $\kappa > 0$,
then slower convergence rates of $I(n)$ would suffice.

%The variances and covariances that appear in the final representations of the variance terms are of the order $O_p(n^{-1})$ by Assumptions~\eqref{ass:mom1}, \eqref{ass:mom21*}, and \eqref{ass:mom21}.
%A comparison of all involved rates reveal that the dominating term in the complete variance expressions are the covariances that appear in the ``second terms''.

%  The \backmatter command formats the subsequent headings so that they
%  are in the journal style.  Please keep this command in your document
%  in this position, right after the final section of the main part of
%  the paper and right before the Acknowledgements, Supporting Information (Supplementary %  Materials),   and References sections.

%\backmatter

%  This section is optional.  Here is where you will want to cite
%  grants, people who helped with the paper, etc.  But keep it short!

%\section*{Acknowledgements}
%The authors thank
%\vspace*{-8pt}

\section*{Acknowledgements}

The authors would like to thank Jan Beyersmann for discussions about the Fine-Gray model and the data set analyzed in this paper.

%  Here, we create the bibliographic entries manually, following the
%  journal style.  If you use this method or use natbib, PLEASE PAY
%  CAREFUL ATTENTION TO THE BIBLIOGRAPHIC STYLE IN A RECENT ISSUE OF
%  THE JOURNAL AND FOLLOW IT!  Failure to follow stylistic conventions
%  just lengthens the time spend copyediting your paper and hence its
%  position in the publication queue should it be accepted.

%  We greatly prefer that you incorporate the references for your
%  article into the body of the article as we have done here
%  (you can use natbib or not as you choose) than use BiBTeX,
%  so that your article is self-contained in one file.
%  If you do use BiBTeX, please use the .bst file that comes with
%  the distribution.  In this case, replace the thebibliography
%  environment below by
%
%  \bibliographystyle{biom}
% \bibliography{mybibilo.bib}

%\bibliographystyle{plainnat}
%\bibliography{references}

%\include{Appendix1/appendix1}

%\addcontentsline{toc}{chapter}{\protect \textbf{References}}
%\include{Bibliography/bibliography}

%\rhead{}
%\lhead{}
%\fancyhead[LE,RO]{\thepage}

%\clearpage
%\phantomsection
%\label{bibliography}
\bibliographystyle{plainnat}
\bibliography{references}
%\markboth{}{}

%\addcontentsline{toc}{chapter}{\protect \textbf{Abstract}}
%\include{Abstract/abstract}
%\addcontentsline{toc}{chapter}{\protect \textbf{References}}
%\include{Bibliography/bibliography}
%\cleardoublepage

%\clearpage
% \cleardoublepage
% \phantomsection
% \addcontentsline{toc}{chapter}{\protect \textbf{Summary}}
%\include{Summary/summary_thesis}
%\cleardoublepage
%\clearpage
%\phantomsection
%\addcontentsline{toc}{chapter}{\protect \textbf{Samenvatting}}
%\include{Samenvatting/samenvatting}
%\addcontentsline{toc}{chapter}{\protect \textbf{Acknowledgement}}
%\include{Acknowledgement/acknowledgement}
%\appendix
%\include{Appendix1/appendix1}
%\include{Appendix2/appendix2}

\end{document}